\documentclass[iop,twocolappendix]{emulateapj-rtx4}

\usepackage{natbib}
\usepackage{mathrsfs}
\usepackage{amssymb}
\usepackage{amsmath}
\newcommand{\secpoint}{\mbox{$''\mskip-7.6mu.\,$}}

\shorttitle{Merging Galaxy Pairs}
\shortauthors{Wilson et al.}

\begin{document}

\title{The MOSDEF Survey: No Significant Enhancement in Star Formation or Deficit in Metallicity in
Merging Galaxy Pairs at $1.5 \lesssim \MakeLowercase{z} 
\lesssim3.5$\altaffilmark{1}}

\author{Tim J. Wilson,\altaffilmark{2}
	Alice E. Shapley,\altaffilmark{3}
	Ryan L. Sanders,\altaffilmark{3}
	Naveen A. Reddy,\altaffilmark{4}
	William R. Freeman,\altaffilmark{4}
	Mariska Kriek,\altaffilmark{5}
	Irene Shivaei,\altaffilmark{6}
	Alison L. Coil,\altaffilmark{7}
	Brian Siana,\altaffilmark{4}
	Bahram Mobasher,\altaffilmark{4}
	Sedona H. Price,\altaffilmark{8}
	Mojegan Azadi,\altaffilmark{9}
	Guillermo Barro,\altaffilmark{10}
	Laura de Groot,\altaffilmark{11}
	Tara Fetherolf,\altaffilmark{4}
	Francesca M. Fornasini,\altaffilmark{9}
	Gene C. K. Leung,\altaffilmark{7}
	Tom O. Zick,\altaffilmark{5}
}

\altaffiltext{1}{Based on data obtained at the W.M. Keck Observatory, which is 
operated as a scientific partnership among the California Institute of Technology, 
the University of California, and the National Aeronautics and Space 
Administration, and was made possible by the generous financial support of the 
W.M. Keck Foundation.}
\altaffiltext{2}{Department of Physics, University of California, Berkeley, 366 
LeConte Hall MC 7300, Berkeley, CA, 94720, USA}
\altaffiltext{3}{Department of Physics and Astronomy, University of California, 
Los Angeles, 430 Portola Plaza, Los Angeles, CA 90095, USA}
\altaffiltext{4}{Department of Physics and Astronomy, University of California, 
Riverside, 900 University Avenue, Riverside, CA 92521, USA}
\altaffiltext{5}{Astronomy Department, University of California at Berkeley, Berkeley, CA 94720, USA}
\altaffiltext{6}{Steward Observatory, University of Arizona, 933 N Cherry Ave, Tucson, AZ 85721, USA}
\altaffiltext{7}{Center for Astrophysics and Space Sciences, Department of Physics, University of California, San Diego, 9500 Gilman Drive., La Jolla, CA 92093, USA}
\altaffiltext{8}{Max-Planck-Institut f\"ur Extraterrestrische Physik, Postfach 1312, Garching, 85741, Germany}
\altaffiltext{9}{Harvard-Smithsonian Center for Astrophysics, 60 Garden Street, Cambridge, MA, 02138, USA}
\altaffiltext{10}{Department of Phyics, University of the Pacific, 3601 Pacific Ave, Stockton, CA 95211, USA}
\altaffiltext{11}{Department of Physics, The College of Wooster, 1189 Beall Avenue, Wooster, OH 44691, USA}

\email{tim.wilson@berkeley.edu}

\begin{abstract}
We study the properties of 30 spectroscopically-identified pairs of galaxies observed during
the peak epoch of star formation in the universe. These systems are drawn from the MOSFIRE Deep
Evolution Field (MOSDEF) Survey at $1.4 \leq z \leq 3.8$, and are interpreted as early-stage
galaxy mergers. Galaxy pairs in our sample are identified as two objects whose spectra were
collected on the same Keck/MOSFIRE spectroscopic slit. Accordingly, all pairs in the sample
have projected separations $R_{{\rm proj}}\leq 60$~kpc. The velocity separation for pairs was
required to be $\Delta v \leq 500 \mbox{ km
s}^{-1}$, which is a standard threshold for defining interacting galaxy pairs at low
redshift. Stellar mass ratios in our sample range from 1.1 to 550, with 12 ratios closer than or
equal to 3:1, the common
definition of a ``major merger." Studies of merging pairs in the local universe indicate an
enhancement in star-formation activity and deficit in gas-phase oxygen abundance relative to
isolated galaxies of the same mass. We compare the MOSDEF pairs sample to 
a control sample of isolated galaxies at the same redshift, finding no measurable SFR enhancement or
metallicity deficit at fixed stellar mass for the pairs sample. The lack of significant
difference between the average properties of pairs and control samples appears in contrast to results from
low-redshift studies, although the small sample size and lower signal-to-noise of the high-redshift
data limit definitive conclusions on redshift evolution. These results
are consistent with some theoretical works suggesting a reduced differential effect of
pre-coalescence mergers on galaxy properties at high redshift -- specifically that
pre-coalescence mergers do not drive strong starbursts.
\end{abstract}

\keywords{galaxies: evolution --- galaxies: interactions --- galaxies: high-redshift}

\section{Introduction}\label{sec:intro}

Galaxies grow in mass through a combination of mergers with other galaxies and 
smooth accretion of baryons and dark matter. Predicting the frequency of both 
major (i.e., with roughly equal masses ) and minor (i.e., with significantly unequal masses) 
mergers as a function of galaxy mass and redshift is therefore an important 
component of hierarchical models of structure formation 
\citep[e.g.,][]{hopkins2010}. At the same time, obtaining empirical constraints on such 
merger rates as a function of galaxy mass and redshift represents a key goal for 
observations of galaxy evolution \citep[e.g.,][]{lotz2011}. In addition to 
quantifying merger rates, both models and observations aim to describe the impact 
of galaxy interactions on the properties of merging and coalesced galaxies.

Simulations of star-forming galaxy mergers predict a characteristic progression of the 
star-formation rate (SFR) throughout the merger event. Relative to the time prior 
to the merger, the SFRs of the merging galaxies are elevated during their extended 
gravitational interaction, and ultimately peak when the galaxies coalesce 
\citep[e.g.,][]{mihos1996,hopkins2008,cox2008}. The degree of enhancement in SFR 
is predicted to depend on galaxy mass ratio. For example, \citet{cox2008} has 
demonstrated that mergers with mass ratio smaller than 3:1 lead to much stronger 
bursts of star formation than mergers with larger mass ratios. Additional factors 
affect the strength of the merger-induced starburst, such as the orientation of 
the orbits of merging galaxies, as well as their structural properties and gas fractions.

In the local universe, the most luminous systems, (i.e., ultraluminous infrared 
galaxies; ULIRGs), appear to be dominated by advanced-stage major mergers during or 
just after coalescence \citep[e.g.][]{sanders1996,tacconi2002}. Pre-coalescence stages of merging at 
$z\sim 0$ have been traced by galaxy pairs. The Sloan Digital Sky Survey (SDSS) 
has yielded a statistical sample of such pairs \citep[e.g.,][]{ellison2008, scudder2012, scudder2015,patton2011,patton2013}, 
identified as galaxies separated 
by both a small projected radius (with upper limits on $R_{{\rm proj}}$ ranging from 30 to 80~kpc) 
and small radial velocity 
difference (with upper limits on $\Delta v$ ranging from $200$ to $500\mbox{ km s}^{-1}$). Members of these 
galaxy pairs are characterized by both enhanced SFRs \citep[e.g., $\sim 60$\% out to 30~kpc;][]{scudder2012} and depressed gas-phase 
oxygen abundances \citep[e.g., $\sim 0.02$~dex;][]{scudder2012} relative to a control sample of isolated galaxies matched in
stellar mass. Such differences are consistent with theoretical models of galaxy mergers in 
which an increase in SFR accompanies the inflow of gas into the central regions of the
merging galaxies, which also tends to dilute the metal content of the interstellar medium
\citep[ISM;][]{hopkins2008,bustamante2018}. A similar enhancement in SFR in galaxy pairs has been
detected out to $z\sim 1$ \citep{lin2007,wong2011}.

\begin{figure*}[t!]
\includegraphics[width=\textwidth]{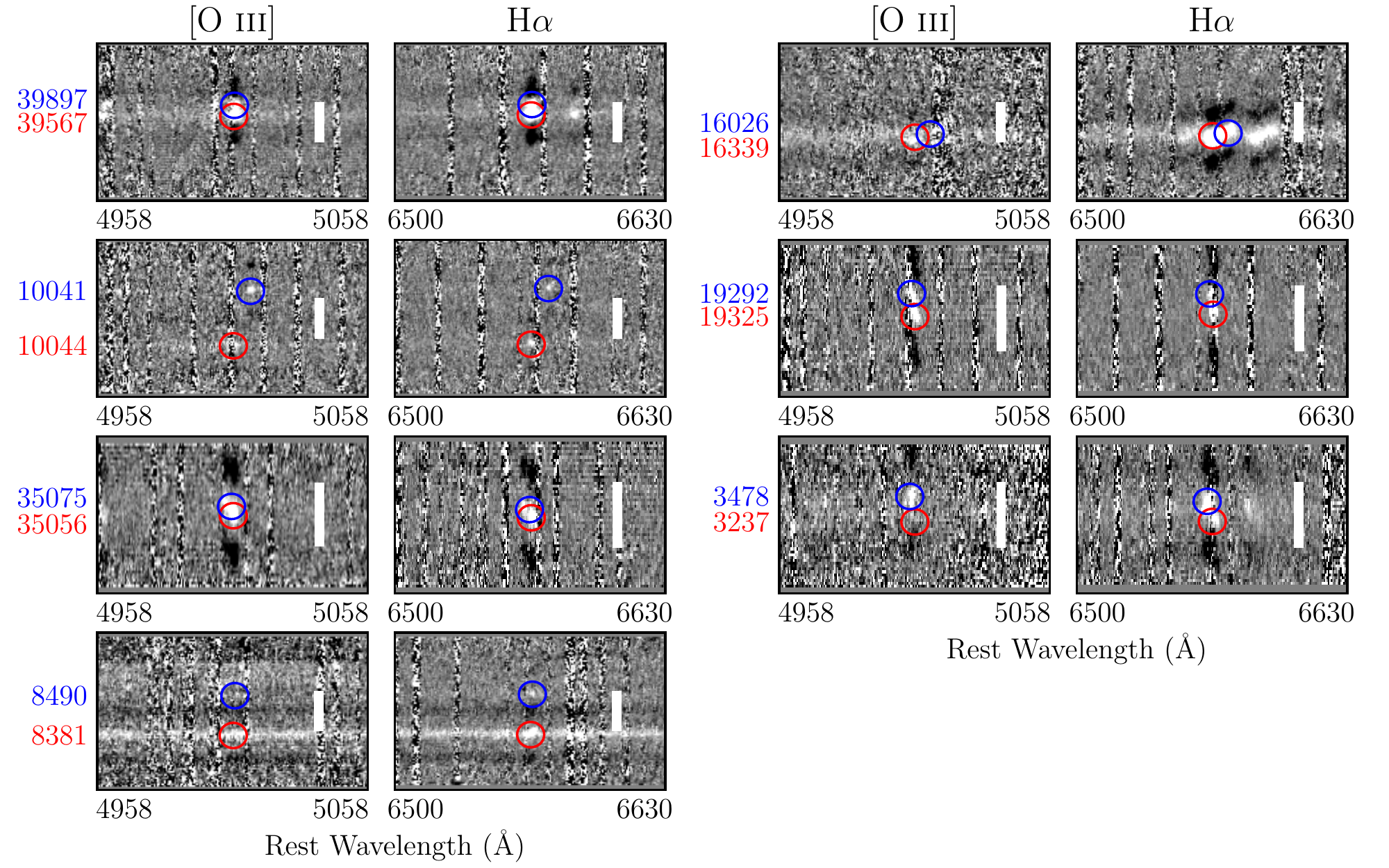}
\caption{Gallery of two-dimensional spectra of galaxy pairs
at $1.37 \leq z \leq 1.70$. Two panels are shown for each galaxy pair, which
zoom in on the rest-frame wavelength ranges centered on [OIII]$\lambda 5007$
and H$\alpha$ for the primary target, and correspond to $\Delta z=\pm 0.01$. At these redshifts [OIII]$\lambda 5007$
falls in the observed $J$ band, while
H$\alpha$ falls in the observed $H$ band. For each
pair, emission from the primary target galaxy is circled in red, while that
from the serendipitous companion is circled in blue. 3D-HST v4.1 catalog
numbers \citep{skelton2014} are given to the left of the [OIII]$\lambda 5007$
panel, with red and blue color-coding corresponding to the circles.
Spectral cut-outs are scaled to the same vertical size on the page
for display purposes,
resulting in a variable angular scale in that dimension. Accordingly, we provide a white vertical
scale bar in each panel indicating the extent of 30 proper kpc, which
corresponds to 3\secpoint65 at the median redshift of the sample.}
\centering
\label{fig:z12d}
\end{figure*}

\begin{figure*}[t!]
\includegraphics[width=0.95\textwidth]{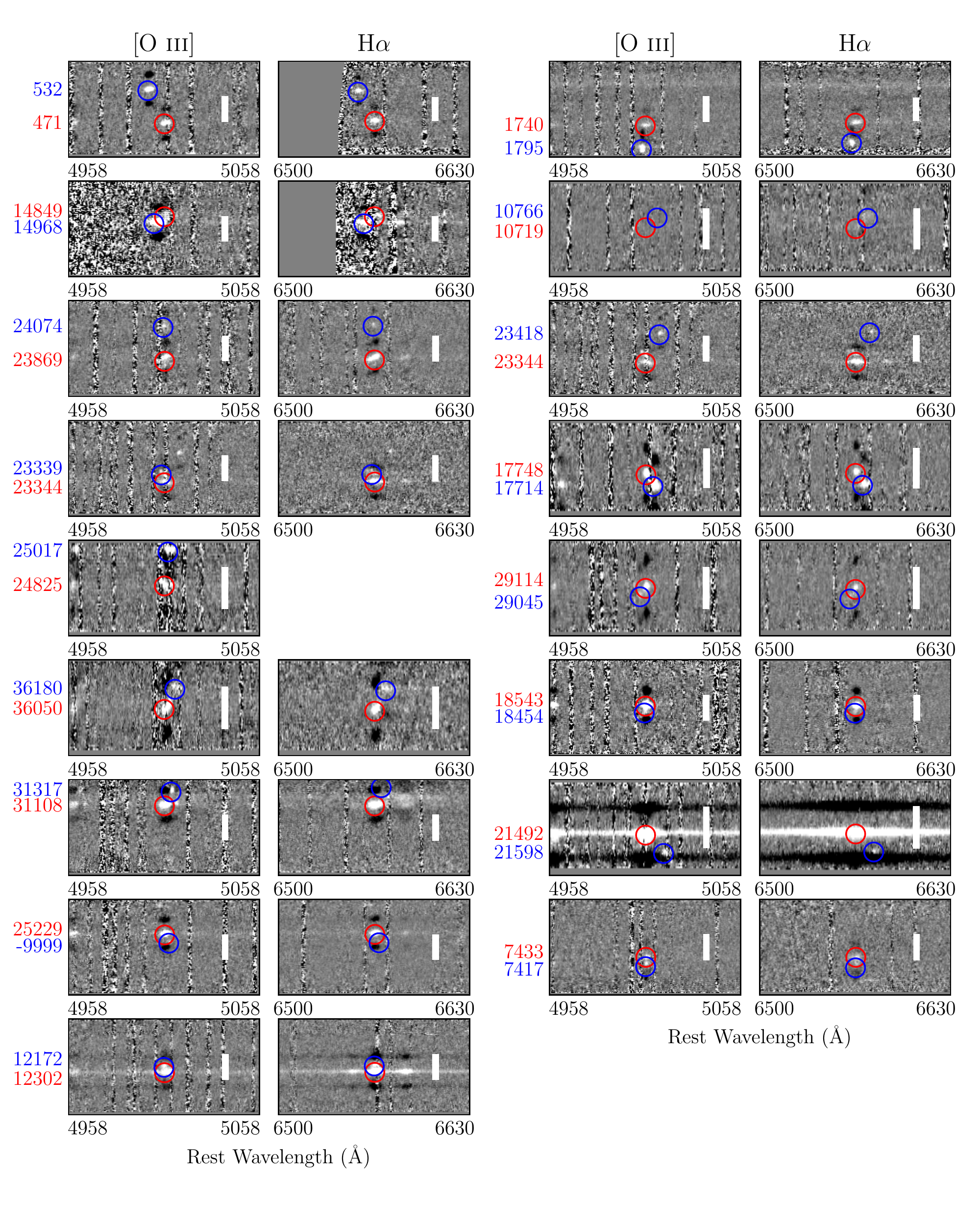}
\caption{Gallery of two-dimensional spectra of galaxy pairs
at $1.90 \leq z \leq 2.70$. Two panels are shown for each galaxy pair, which
zoom in on the rest-frame wavelength ranges centered on [OIII]$\lambda 5007$
and H$\alpha$ for the primary target, and correspond to $\Delta z=\pm 0.01$. At these redshifts [OIII]$\lambda 5007$
falls in the observed $H$ band, while
H$\alpha$ falls in the observed $K$ band. All symbols, labels, and scale bars as in Figure~\ref{fig:z12d}.
The $z=2.3$ pair GOODSN-24825/25017 lacks coverage of H$\alpha$ because it was observed as a filler target
on a MOSDEF mask targeting $1.37\leq z \leq 1.70$ galaxies and therefore not observed in the $K$ band.
}
\centering
\label{fig:z22d}
\end{figure*}

\begin{figure*}[t!]
\includegraphics[width=\textwidth]{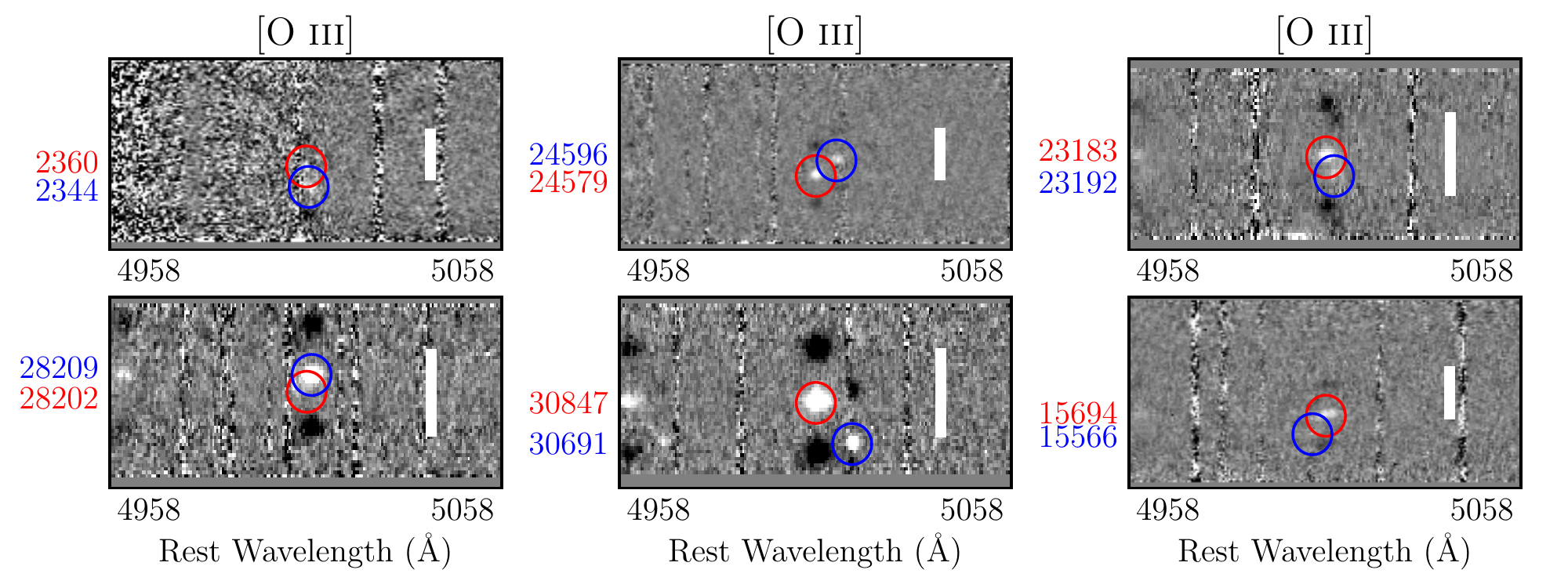}
\caption{Gallery of two-dimensional spectra of galaxy pairs
at $2.95 \leq z \leq 3.80$. One panel is shown for each galaxy pair, which
zooms in on the rest-frame wavelength ranges centered on [OIII]$\lambda 5007$ for the primary target,
and corresponds to $\Delta z=\pm 0.01$. At these redshifts [OIII]$\lambda 5007$
falls in the observed $K$ band.  All symbols, labels, and scale bars as in Figure~\ref{fig:z12d}.
}
\centering
\label{fig:z32d}
\end{figure*}

Galaxy mergers have now been identified out to $z\sim 6$ \citep{ventou2017}. In the early universe 
(i.e., at $z>1$), primarily one of two techniques is employed to flag merging 
systems. First, it is common to use morphological signatures to identify ongoing 
or recently completed merger events. Galaxies have been visually classified as 
mergers on the basis of morphological features such as tidal tails and bridges, 
and double nuclei \citep{lofthouse2017}, and also identified as interacting based 
on non-parametric morphological statistics such as the Gini and $M_{20}$ 
coefficients \citep{lotz2004}, or the concentration (C), asymmetry (A), and 
clumpiness (S) statistics \citep{conselice2014}. The second technique for flagging 
mergers is through galaxy pairs. Many studies aiming to quantify the merger 
fraction and rate at $z>1$ have been based on photometric pairs, which consist of 
galaxies within a small projected radius and small difference in photometric 
redshift \citep[e.g.,][]{man2012,man2016,mantha2018,williams2011}. In some cases 
\citep{bluck2009}, the photometric redshift for only one of the galaxies is known.  
The possibility of contamination by chance projections must therefore be accounted 
for, especially when the redshift of a potential companion galaxy is unknown. 
Recently, merging pairs at $z>1$ have also been identified spectroscopically, 
based on rest-frame ultraviolet spectra \citep{tasca2014,ventou2017}. However, the 
sensitivity of rest-frame UV features to large-scale galaxy outflows 
\citep{pettini2001,shapley2003,steidel2010} limits the accuracy with which such 
galaxy systemic redshifts, and therefore merger dynamics, can be measured.

To date, most studies of merging pairs at $z>1$ have focused on global statistics 
such as the merger fraction and rate, as opposed to systematic studies of 
the impact of close interactions on the properties of merging galaxies.  In this 
work, we focus on the latter, based on a sample of galaxy mergers identified at $1.4 \leq z \leq 3.8$ within the MOSFIRE Deep 
Evolution Field (MOSDEF) survey \citep{kriek2015}. The extensive rest-optical spectroscopic
coverage of the MOSDEF survey enables us to assemble a clean sample of galaxy
pairs that are not only close on the sky but also in redshift space. With spectroscopic pairs, there
is little possibility of contamination by chance projections of completely unassociated galaxies.
Such chance projections can arise when pairs are identified on the basis of photometric redshifts, 
given their associated uncertainties at high redshift.\footnote{We note that proximity in redshift
space does not guarantee merging, as galaxy pairs offset by tens of proper kpc in $R_{proj}$
and with line-of-sight velocity separations of up a few hundreds of $\mbox{ km s}^{-1}$
may not be bound and destined to merge \citep{moreno2013}. However, with spectroscopic redshift measurements,
we can at least apply the same proximity criteria in velocity space that is used for studies
of local galaxy pairs.} Furthermore, we have estimated 
key galaxy properties such as SFR, stellar mass ($M_*$), and gas-phase oxygen 
abundance for both merging and isolated systems, and can therefore study for the 
first time the effect of interactions on star-formation activity and chemical 
enrichment in distant star-forming galaxies.

In Section~\ref{sec:mosdef}, we present the details of the MOSDEF survey and the galaxy 
properties analyzed in this work. Section~\ref{sec:pairs} discusses the selection and
properties of spectroscopically-determined merging pairs in MOSDEF, while 
Section~\ref{sec:control} describes the selection of our control sample of isolated 
galaxies used for systematic comparison with mergers. In Section~\ref{sec:properties}, we 
investigate the effect of mergers on star formation and metal enrichment through 
analysis of the SFR-$M_*$ main sequence and stellar mass-metallicity relation 
(MZR) for both merging pairs and isolated control galaxies. We present a 
discussion of our results and describe future work in Section \ref{sec:discussion}. 
Throughout this paper, we adopt cosmological parameters of $H_0=70 \mbox{ km 
s}^{-1} \mbox{ Mpc}^{-1}$, $\Omega_M = 0.30$, and $\Omega_{\Lambda}=0.7$.

\section{The MOSDEF Survey}\label{sec:mosdef}
We assembled a sample of spectroscopically-confirmed merging pairs from the MOSDEF 
survey. With MOSDEF, we performed a large survey of the rest-frame optical spectra 
of $\sim 1500$ galaxies spanning $1.4 \leq z \leq 3.8$. Spectra were collected for 
MOSDEF galaxy targets using the MOSFIRE spectrograph \citep{mclean2012} on the 
Keck~I telescope. For a full description of the MOSDEF survey observations and 
data reduction, we refer readers to \citet{kriek2015}. Here we provide the survey 
information most relevant to the current work.

MOSDEF observing runs comprised 48.5 MOSFIRE nights between 2012 December and 2016 
May. Galaxies in the MOSDEF sample are concentrated in three redshift intervals where 
strong rest-optical emission lines fall within windows of atmospheric transmission 
$(1.37 \le z \le 1.70,2.09 \le z \le 2.61, \mbox{ and } 2.95 \le z \le 3.80)$. The targeted 
galaxies were selected from the photometric and spectroscopic catalogs constructed
as part of the 3D-HST survey \citep{brammer2012,momcheva2016,skelton2014} down to limiting 
{\it HST}/WFC3 F160W AB magnitudes of 24.0, 24.5, and 25.0, respectively at $1.37 \le 
z \le 1.70$, $2.09 \le z \le 2.61$, and $2.95 \le z \le 3.80$. As these galaxies 
are primarily located in three CANDELS fields: AEGIS, COSMOS, and GOODSN 
\citep{grogin2011,koekemoer2011}, they have extensive multi-wavelength photometric 
coverage \citep{skelton2014} from which stellar population parameters and photometric 
redshifts are derived.

Targets at $1.37 \le z \le 1.70$ were typically observed using 1-hour exposures in 
the $Y$, $J$, and $H$ bands; those at $2.09 \le z \le 2.61$ were observed using 2-hour 
exposures in the $J$, $H$, and $K$ bands; and those at $2.95 \le z \le 3.80$ were 
observed using 2-hour exposures in the $H$ and $K$ bands. Our MOSFIRE multi-object 
slitmasks typically contained $\sim 30$ 0\secpoint7 slits, yielding a 
resolution of 3400 in Y, 3300 in J, 3650 and H, and 3600 in $K$. In practice, galaxy 
pairs are identified in this work on the basis of the strongest rest-frame optical 
features, which are H$\alpha$ and [OIII]$\lambda 5007$. H$\alpha$ is measured in 
the $H$ and $K$ bands, respectively, at $1.37 \le z \le 1.70$ and $1.90 \le z \le 
2.61$, while [OIII]$\lambda 5007$ is measured in the $J$ and $H$ bands, respectively,
over the same redshift ranges and in the $K$ band at $2.95 \le z \le 3.80$.

We used a custom IDL pipeline to reduce the raw data and produce two-dimensional 
spectra in each filter, as described in \citet{kriek2015}. One-dimensional science and 
error spectra were then optimally extracted from the two-dimensional spectra \citep{freeman2017}. The 
final MOSDEF sample contains 1493 {\it primary} targets, 66 of which represent 
duplicate observations. In addition, the sample includes 165 galaxies that 
serendipitously fell within MOSDEF slits and for which we measured spectroscopic 
redshifts (hereafter {\it serendips}). 

Due to the nature of our MOSFIRE slit observations and the manner in which serendips
were identified, special care is required to obtain accurate flux and wavelength 
information for serendips. The coordinates of MOSDEF primary targets 
determined the location of each MOSFIRE slit, and therefore primary targets are 
well centered in the slits. We identify galaxy pairs by the presence of a serendip 
companion galaxy whose light also falls in the slit of the primary target, and 
whose spectrum yields a redshift close to that of the primary target. As the 
position of the serendip was not taken into account when designing MOSDEF slit 
masks, such galaxies are not necessarily centered across the slits that capture 
their light. When we apply slit-loss corrections to the spectra for each galaxy in 
each filter \citep{kriek2015}, we take into account the potentially off-center nature 
of serendip sources. The potential horizontal offset of a serendip also leads to a small 
offset in the actual wavelength solution that should be calculated for the 
serendip relative to what is derived based on sky lines that fill and are centered 
in the slit. However, the velocity offsets corresponding to spatial offsets of 
less than or equal to half of a slit width are small ($\leq 50 \mbox{ km s}^{-1}$) 
compared to the range of velocity offsets between primary and serendip objects 
considered in this work, and do not affect any of our conclusions. Therefore, 
we do not correct for such offsets.

We measured emission-line fluxes by fitting Gaussian profiles to one-dimensional 
spectra. The MOSFIRE redshift for each galaxy, $z_{\rm{MOSFIRE}}$, was estimated 
from the centroid of the highest S/N feature detected (i.e., typically H$\alpha$ 
or [OIII]$\lambda5007$). Balmer emission-line fluxes were corrected for underlying 
stellar absorption based on the best-fit stellar population model to the observed 
broadband spectral energy distribution (SED). Several galaxy properties were 
derived for our targets based on MOSFIRE emission-line fluxes and 
existing multi-wavelength imaging data. These include the nebular extinction, 
$E(B-V)_{{\rm neb}}$, based on the observed H$\alpha$/H$\beta$ Balmer decrement and 
assuming a Milky Way dust extinction curve \citep{cardelli1989}. H$\alpha$ SFRs  (SFR(H$\alpha$))
were then estimated from the dust-corrected H$\alpha$ luminosities, based on the 
\citet{hao2011} update to the calibration of \citet{kennicutt1998} 
and assuming a \citet{chabrier2003} IMF. 
Stellar masses ($M_*$) were estimated by using the fitting program, FAST \citep{kriek2009}, 
to fit the stellar population synthesis models of \citet{conroy2009} to galaxy 
broadband photometric SEDs, assuming a \citet{chabrier2003} IMF, solar metallicity, and a 
\citet{calzetti2000} dust attenuation law. We also assumed delayed exponential
star-formation histories of the form $\mbox{SFR}\propto t \exp(-t/\tau)$,
with $t$ the time since star formation commenced and $\tau$ the star formation
decay timescale.  Accordingly, SED fitting also yielded an independent 
estimate of the SFR, i.e., SFR(SED). Finally, gas-phase oxygen abundances were
estimated from two empirical calibrations commonly applied at high redshift. We 
used the calibrations of \citet{pettinipagel2004} based on the $N2$ and $O3N2$ 
emission-line indicators, which are defined as, $N2 
=\mbox{[NII]}\lambda6584/\mbox{H}\alpha$ and $O3N2 = 
(\mbox{[OIII]}\lambda5007/\mbox{H}\beta)/(\mbox{[NII]}/\mbox{H}\alpha)$.
These calibrations are:

\begin{equation}\label{eq:n2}
12+\log(\mbox{O/H})_{\rm{N2}} = 8.90 + 0.57 \times \log(N2)
\end{equation}

\noindent and

\begin{equation}\label{eq:o3n2}
12+\log(\mbox{O/H})_{\rm{O3N2}} = 8.73 + 0.32 \times \log(O3N2)
\end{equation}

Although there has been considerable debate in the literature
regarding the validity of these locally-calibrated metallicity indicators in an {\it absolute} sense for 
high-redshift galaxies \citep[e.g.,][]{sanders2015,steidel2014}, these indicators should be
adequate for estimating {\it relative} metallicity differences among high-redshift galaxies.

\begin{figure*}[t!]
\includegraphics[width=\textwidth]{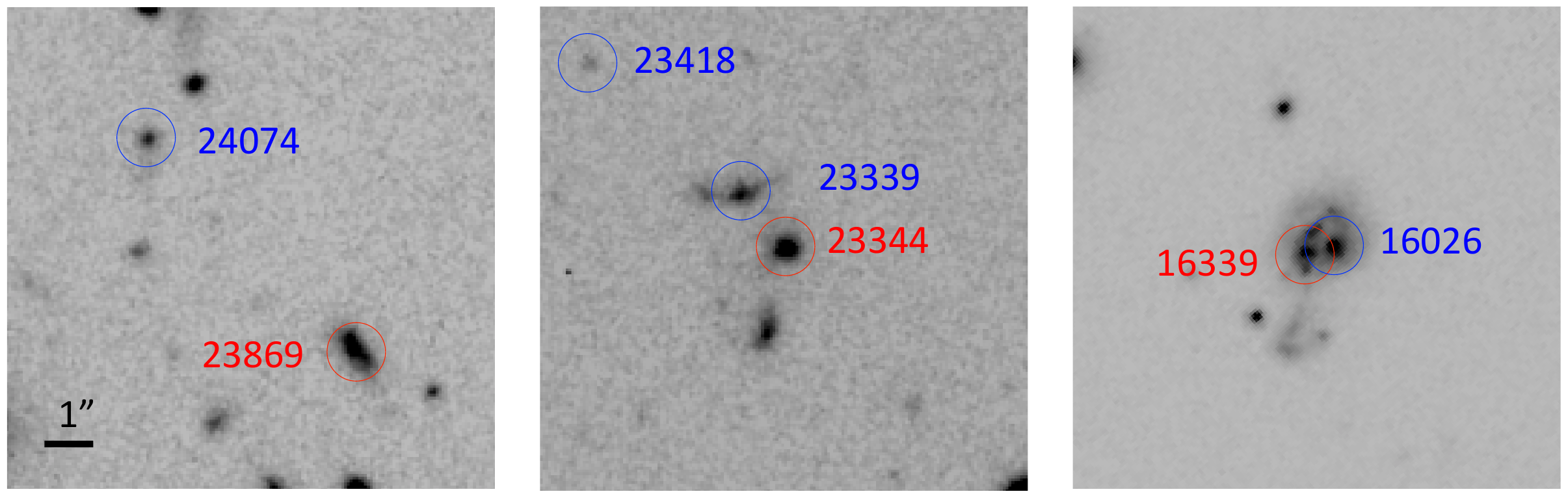}
\caption{{\it HST} WFC3/F160W postage stamps for pairs in our sample, demonstrating
the variety of systems covered. Each postage stamp is 10"$\times$10",
oriented with North up and East to the left. In each image, the primary
target is indicated in red, while the serendips are labeled in blue. Further details
of each merging pair are provided in Tables~\ref{tab:basic} and ~\ref{tab:physical}.
{\bf Left}: A wide-separation pair, consisting of GOODSN-23869
(primary; $z=2.2438$) and GOODSN-24074 (serendip; $z=2.2433$).
{\bf Center}: A triple of associated galaxies, including GOODSN-23344
(primary; $z=2.4839$), GOODSN-23339 (serendip; $z=2.4828$), and
GOODSN-23418 (serendip; $z=2.4889$).  The galaxy just to the south
of GOODSN-23344 (i.e., GOODSN-23271) does not have a spectroscopic redshift,
but its photometric redshift is consistent with the spectroscopic redshift
of GOODSN-23344. {\bf Right}: One of the pairs
with the smallest observed separation, suggesting coalescence. This
pair consists of AEGIS-16339 (primary; $z=1.5291$) and AEGIS-16026
(serendip; $z=1.5320$). This system also shows qualitative
evidence for extended tidal features.}
\centering
\label{fig:postage}
\end{figure*}

\section{Merging pair selection}\label{sec:pairs}
In order to identify spectroscopic merging pairs within MOSDEF, 
we applied the following criteria, which are broadly motivated by the low-redshift study of \citet{ellison2008}:
\begin{enumerate}
        \item The spectra of two or more galaxies must have been collected in a single MOSDEF spectroscopic slit, comprising the primary galaxy target and at least one serendip. This criterion effectively translates into a cut on $R_{{\rm proj}}$, given the typical size of MOSFIRE slits and the small variation in angular size for a fixed proper distance over the redshift range of our sample.
        \item The primary and serendip galaxies must both have secure MOSFIRE spectroscopic redshifts.
        \item The two objects must be separated by \\ $\Delta v = c | z_{{\rm primary}} - z_{{\rm serendip}} |/(1+z_{{\rm primary}}) \leq 500 \mbox{ km s}^{-1}$,
where $z_{{\rm primary}}$ is the redshift of the primary target and $z_{{\rm serendip}}$ is the redshift of the serendip.
\end{enumerate}

Adopting the above criteria, we have spectroscopically identified 31
merging pairs, one of which was observed twice (with primary and 
serendip classifications reversed). Given the small sample size, we
carefully inspected each pair in both two-dimensional spectra and {\it HST}
F160W images to confirm the validity of our pair identifications. In particular,
for pairs with small $R_{{\rm proj}}$, we wished to check that two distinct
emission lines could be ascertained in the two-dimensional spectra, and
that two distinct brightness concentrations could be determined in the {\it
HST} images, corresponding to the separate 3D-HST catalog identifiers. This close analysis caused us to remove one ``pair" from our
initial sample, as the {\it HST} image revealed the serendip object to be a
single bright knot within the more extended light distribution of the
primary. Furthermore, the two-dimensional MOSFIRE spectrum showed an extended,
tilted emission line that did not clearly break up into two components.
Accordingly, our final sample consists of 30 dynamical pairs.

One of the 30 serendips (primary target COSMOS-25229, $z=2.1813$) was
clearly apparent in the {\it HST} F160W image, but had no identifier or multi-wavelength SED in the
3D-HST photometric catalog \citep{skelton2014}. We include the
corresponding pair in our analysis, as the properties of the primary target
can still be considered differentially with those of our control sample
described in the next section. We also note that 15 of the serendips were
contained within the MOSDEF parent catalog, and could have been targeted
for spectroscopy as part of the MOSDEF survey, whereas 15 were fainter than
the MOSDEF limits for targeted spectroscopy or had photometric redshifts outside the MOSDEF target redshift
ranges. There is also one case in which {\it two} associated serendips
were identified on the slit along with the primary target (i.e.,
primary GOODSN-23344, $z=2.4839$). Accordingly, our 30 pairs comprise 29 primary galaxies
and 30 serendips for a total of 59 galaxies.

Galleries of two-dimensional [OIII] and H$\alpha$ emission-line spectra indicating
primary and serendip objects are contained in Figures~\ref{fig:z12d}--\ref{fig:z32d}. For the subsample
of $z\sim 3$ pairs, we only show [OIII], since these pairs lack H$\alpha$
coverage. For pairs at $z\leq 2.7$ with both [OIII] and H$\alpha$
coverage, it can be seen that the two-dimensional emission-line morphologies are typically
very similar in [OIII] and H$\alpha$. Key properties of our galaxy pairs
are summarized in Tables~\ref{tab:basic} (redshift, apparent magnitude,
projected physical separation, and line-of-sight velocity separation) and
\ref{tab:physical} (stellar mass, SFR, and metallicity). In Figure~\ref{fig:postage}, we show example {\it HST}
F160W postage stamps indicating the range of properties in our sample: a
widely separated pair; a triple of associated galaxies; and a small-separation pair
that is apparently close to coalescence.

The 30 pairs in our sample span the redshift interval $1.4 \leq z \leq
3.5$. Since the pairs at $z>2.65$ lack coverage of H$\alpha$ emission, we
are unable to determine robust Balmer decrements and therefore
dust-corrected SFRs based on Balmer emission lines. For these six $z\sim 3$
pairs, we also do not have access to the $N2$ and $O3N2$ metallicity
indicators \citep{pettinipagel2004}, which are commonly applied at $z\leq
2.65$. Accordingly, we list the basic parameters for the $z\sim 3$ pairs,
but do not analyze their differential physical properties in the spaces of
SFR, mass, and metallicity in Section~\ref{sec:properties}. Our
differential analysis focuses on the 24 remaining pairs at $1.4 \leq z \leq
2.6$. This redshift range overlaps the epoch of peak star formation in
the history of the universe \citep{md14}. Therefore, our study enables us
to trace the impact of early-stage mergers on galaxy properties during its
most active period. In order to maximize our sample size, we analyze all 24
pairs together. Although the sample spans a significant range in cosmic
time (1.9~Gyr), we justify the joint analysis across this redshift interval
based on the results of \citet{shivaei2015}, who demonstrated that there is no
significant evolution in the star-forming main sequence (i.e., the SFR vs.
stellar mass relation) between $z \sim 1.4$ and $2.6$.

\begin{figure*}[t!]
\includegraphics[width=\textwidth]{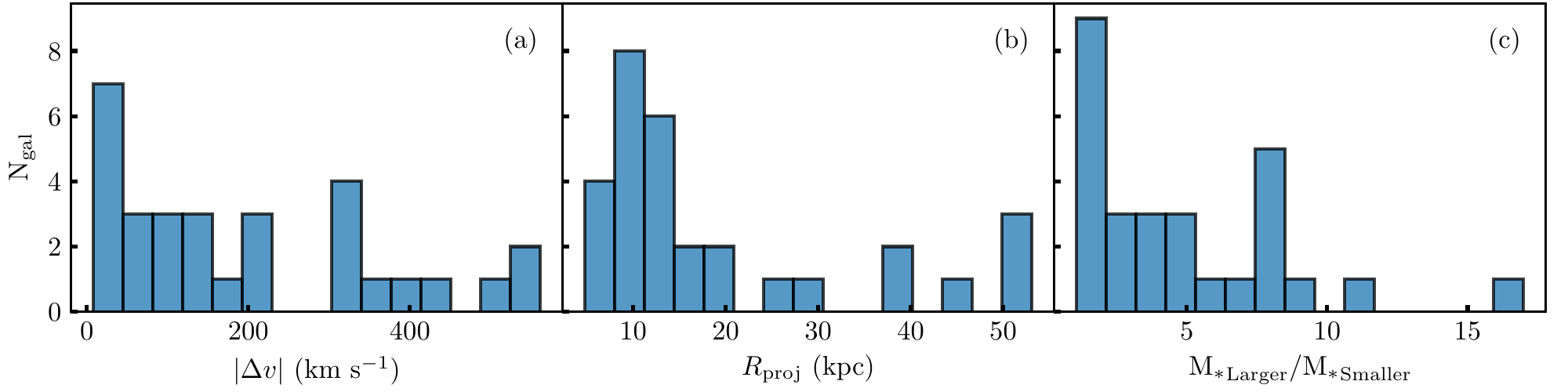}
\caption{{\bf (a)}: Histogram of
line-of-sight velocity difference, $|\Delta v |$ between primary
and serendip objects. {\bf (b)}: Histogram of projected physical separation
$R_{{\rm proj}}$ between primary and serendip objects. {\bf (c)}: Histogram of
stellar mass ratio between more massive and less massive objects,
irrespective of which galaxy is the ``primary" and which is the
``serendip." For improved display quality, we exclude from this panel the one pair with a mass
ratio of 550, zooming on the pairs ranging in mass ratio between 1.1 and 17.}
\centering
\label{fig:hist}
\end{figure*}

One of the unique features of our sample is the precise spectroscopic redshifts
available for our targets. We use these redshifts to calculate the velocity
separations between primary objects and serendips, and adopt an upper
limit for pair velocity separations of $\Delta v = 500 \mbox { km s}^{-1}$.
This limit is chosen to match the one adopted in  \citet{ellison2008}
for analysis of merging pairs in the local universe. We cannot establish definitively
that the spectroscopic pairs we have identified will merge. However, simulations show that over this
range of velocity separations, the majority of pairs are bound and will eventually coalesce \citep[e.g.,][]{moreno2013}.
The distribution of line-of-sight velocity separations is shown in Figure~\ref{fig:hist}a.

We also calculated the projected physical separation, $R_{{\rm proj}}$, between
primary and serendip objects using their mean redshift and sky coordinates
in the 3D-HST catalog \citep{skelton2014}. In the one case of the serendip
unidentified in the 3D-HST catalog, we measured its position directly from
the {\it HST} F160W image. The distribution in $R_{{\rm proj}}$ is shown in
Figure~\ref{fig:hist}b. In part due to the typical MOSFIRE slit length of
7\secpoint1, all of the projected physical separations in our sample are
smaller than 60~kpc, which is well within the limit of 80~kpc adopted by
\citet{ellison2008}. As another basic pair parameter, we estimated the stellar
mass ratio for each pair, normalized in each case as the ratio between the
mass of the more massive galaxy to that of the less massive galaxy (as
opposed to the primary-to-serendip mass ratio). The distribution of mass
ratios in our sample is shown in Figure~\ref{fig:hist}c. This distribution
spans from 1.1 up to 550, but all except one of the pairs have mass ratios
that fall between 1.1 and 17 (this smaller range in mass ratio is shown
in Figure~\ref{fig:hist}c, for improved display quality). Out of our pair sample, 12 have mass ratios
of 3:1 or smaller, which would cause them to be classified as ``major mergers''
\citep{cox2008}.

One potential limitation of our dataset of merging pairs is its incompleteness due to the small fraction
of on-sky area surrounding each primary galaxy that is sampled by the MOSFIRE spectroscopic slit.
Over the range of redshifts of our pairs sample, the typical MOSFIRE slit dimensions of 7\secpoint1$\times$0\secpoint7
subtend only $\sim 3$\% of the on-sky area out to a projected physical radius of 60~kpc. The small fractional
area subtended by the MOSFIRE slit may have caused us to miss companion galaxies that are actually closer to the primary
galaxy in projected radius than the companions we identified --  and therefore more relevant from the standpoint
of merger interactions -- but do not happen to fall inside the MOSFIRE slit. 

In order to assess the importance of this 
effect, we searched within a projected radius of 60~kpc of each primary galaxy for any additional galaxies with spectroscopic
redshifts placing them within $500 \mbox{ km s}^{-1}$ of the primary galaxy's spectroscopic redshift.
If no spectroscopic redshift was available for galaxies within 60~kpc, we considered
the ``grism" redshifts from the 3D-HST catalogs based on a
combined fit to the broadband photometry and 3D-HST grism spectra,
$z_{{\rm grism}}$.  If no $z_{{\rm grism}}$ was available, we used the 3D-HST
photometric redshift, $z_{{\rm phot}}$, determined from the broadband
photometry \citep{momcheva2016}. For neighboring galaxies with only grism or photometric redshifts,
we identified a galaxy as a potential companion if the 68\% confidence interval of its redshift probability
distribution encompassed the spectroscopic redshift of the primary galaxy.
According to these criteria, we found that 21 out of 29 primary galaxies in our sample
had additional potential galaxy companions within a projected radius of 60~kpc, including only
a single new companion that was spectroscopically confirmed. The remaining  potential companion galaxies
had only photometric or grism redshifts, whose uncertainties are large enough to preclude establishing a real
physical association. Furthermore, in only 3 cases was the new companion galaxy at a smaller
projected radius from the primary galaxy than the companion we had discovered within the MOSFIRE slit.
Accordingly, we have demonstrated that our set of galaxy pairs is not missing a significant number
of additional companions that fall outside the MOSFIRE slits but are at smaller separations from our primary
galaxies than the companions identified within the MOSFIRE slits.

\section{Control group selection}\label{sec:control}

In order to establish the differential effects of merging on galaxy
properties, we identified a control sample of
spectroscopically-confirmed galaxies within the MOSDEF survey that
are characterized by similar galaxy properties (e.g., $M_*$, $z$) to those of our
pairs, yet are confirmed to be lacking a physically associated
companion galaxy. In other words, MOSDEF galaxies in the control
sample are isolated. In assembling this control sample, we must
bear in mind the fact that the MOSDEF pairs sample is highly
incomplete. Indeed, given the small fraction of 360$^\circ$
subtended by the MOSFIRE slit, there remains the possibility that a
physically associated galaxy may be present within 60~kpc
(projected) of one of the MOSDEF primary targets (i.e., the
separation that encompasses all of our MOSDEF pairs) but simply
doesn't fall within the MOSFIRE slit. In such cases, the spectrum
of the neighboring galaxy would not be collected, and this pair of
galaxies would not be included in the MOSDEF pairs sample as
defined in Section~\ref{sec:pairs}. In order to determine a sample
of truly isolated galaxies within the set of MOSDEF primary
targets, we considered each primary target in turn, searching
through the 3D-HST photometry and redshift catalogs to identify
possible neighbors that would not be flagged by our
MOSFIRE-slit-based pairs criteria.

Based on the distribution of $R_{{\rm proj}}$ in the MOSDEF pairs sample,
we defined 60 projected kpc as the radius within which to search
for companion galaxies around MOSDEF primary targets. In practice,
we calculated the corresponding angular radius on the sky, given
the redshift of the primary target. We then identified any possible
companion galaxies in the 3D-HST survey whose coordinates placed
them within this angular radius. For each possible companion, we
estimated $\Delta z = | z_{{\rm primary}} - z_{{\rm companion}} |$, where
$z_{{\rm companion}}$ is the redshift of the potential companion galaxy.
In order to determine if the apparent companion based on angular
separation was also potentially associated with each primary target
in redshift space, we applied criteria that depended on the nature
of the redshift measurement for the apparent companion. If the
apparent companion had a previously-measured spectroscopic redshift
compiled by the 3D-HST survey \citep{brammer2012,momcheva2016} or one measured in
our own MOSDEF spectroscopy, we classified it
as a true companion if $|z_{{\rm primary}} - z_{{\rm companion}}| < 0.01$. If
no spectroscopic redshift was available for the apparent companion,
we used the 3D-HST grism redshifts ($z_{{\rm grism}}$). 
If no $z_{{\rm grism}}$ was available, we used the 3D-HST
photometric redshift, $z_{{\rm phot}}$ \citep{momcheva2016}. In the case of grism and photometric
redshifts, we classified the apparent companion as a
true companion if $z_{primary}$ was within the 68\% confidence interval
of the grism or photometric redshift probability distribution.

\begin{figure*}[t!]
\includegraphics[width=\textwidth]{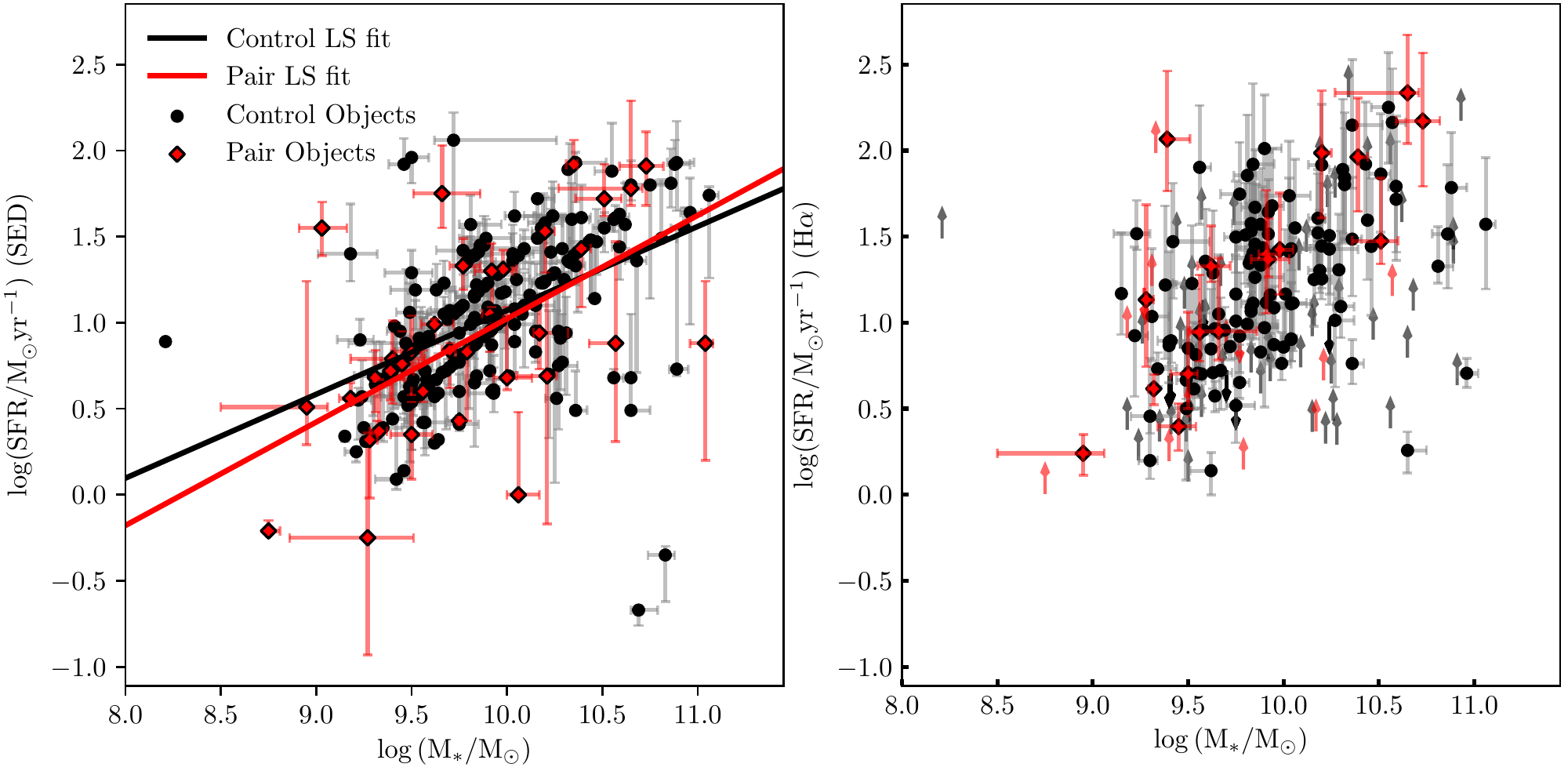}
\caption{SFR-$M_*$ relation for pairs and control objects at $1.4 \leq z \leq 2.6$.
Both primary and serendip objects are plotted for the pairs sample.
{\bf Left:} In this panel, SFR(SED) is estimated from the best-fit stellar population model
to the broadband SED. Galaxies in pairs are indicated with red symbols, while
control sample galaxies are shown with black symbols. Galaxies identified
as AGNs are not plotted, along with those identified as ``quiescent" (i.e., $\log(M_*)>11$ and $\log(\mbox{SFR(SED)})<0.2$,
or $\log(\mbox{SFR(SED)})<-1$).
An ordinary least-squares regression was performed for both pairs and control galaxies (red
and black lines, respectively), yielding very similar fits, which are consistent
within the errors.  {\bf Right:} In this panel, SFR(H$\alpha$) is estimated
from dust-corrected H$\alpha$ luminosity. Our samples of SFR(H$\alpha$) measurements
for both pair and control objects include a significant fraction of limits, predominantly
lower limits resulting from H$\beta$ upper limit. Symbols are as in the left-hand
panel, and AGNs and quiescent objects are removed for this panel as well. Limits for pairs (controls) are
indicated with red (grey) arrows. Given the significant presence of limits in our
SFR(H$\alpha$) samples, we do not perform actual regressions on the SFR(H$\alpha$) vs. $M_*$
distributions for pairs and controls, but we note their overall similarity.}
\centering
\label{fig:sfrm}
\end{figure*}

Accordingly, we identified isolated control galaxies as those that
had no angular neighbors within 60~kpc projected on the sky that
were also true companions in redshift space (i.e., with
$|z_{{\rm primary}} - z_{{\rm companion}}|<0.01$ for apparent companions with
existing $z_{{\rm spec}}$, or $z_{{\rm primary}}$ within the 68\% redshift
confidence interval of apparent companions with existing
$z_{{\rm grism}}$ or $z_{{\rm phot}}$). Using a more conservative isolation criterion
for grism and photometric redshifts (i.e., forcing $z_{{\rm primary}}$ to lie
outside the 95\% grism or photometric redshift confidence interval) yields a 
control sample that is more than two times smaller, 
but results that are statistically consistent with those presented in Section~\ref{sec:properties}. 
Even with the $\sim 1\sigma$ threshold in $|z_{{\rm primary}} - z_{{\rm companion}}|$ when $z_{{\rm companion}}$
was based on a $z_{{\rm grism}}$ or $z_{{\rm phot}}$, we likely excluded a large number of isolated galaxies from the control
sample that were not actually in physically associated pairs. 
Our requirements for isolation yield a sample of 372 MOSDEF galaxies,
242 of which are at $1.4 \leq z \leq 2.6$, i.e., at redshifts where we can measure H$\alpha$ and
[NII]$\lambda 6584$, in addition to H$\beta$ and [OIII]$\lambda 5007$, and therefore obtain measurements
of physical properties such as SFR(H$\alpha$) and metallicity. These 242 galaxies in our control sample 
have a mean [median] redshift of $z=2.15\pm0.02$ [$z=2.24\pm0.02$] and a mean [median] stellar mass of $\log(M_*/M_{\odot})=10.05\pm0.04$ 
[$\log(M_*/M_{\odot})=9.92\pm0.04$], as compared with a mean [median] redshift
and stellar mass of $z=2.07\pm0.05$ [$z=2.23\pm0.06$] and  $\log(M_*/M_{\odot})=9.85 \pm 0.10$ [$\log(M_*/M_{\odot})=9.77 \pm 0.07$], respectively, for our pairs sample at $z<3$.  
When relevant in the following section, we consider a subset of the control sample that is even more precisely matched
in median stellar mass to the pairs sample. Some local studies have also constructed control samples matched in environmental density \citep[e.g.,][]{ellison2010,
alonso2012}. Given that we do not have robust environmental measures for the MOSDEF sample, we limit our matching
to redshift and stellar mass. We also note that previous work has matched pair and control galaxies
on a galaxy-by-galaxy basis \citep[e.g.,][]{ellison2008, scudder2012}. In our analysis, however, our
control sample of isolated galaxies is matched as a whole
to have the same typical redshift and stellar mass as that of our significantly smaller pairs sample.
In summary, the galaxies in our control sample
lack physically associated companions and therefore constitute
a powerful comparison dataset alongside our sample of galaxy pairs within the same redshift range.

\section{Physical Properties of High-Redshift Galaxy Pairs}\label{sec:properties}

Studies of galaxy pairs in the local universe indicate
measurable differences in their star-formation activity
and gas-phase oxygen abundances compared to isolated
galaxies of similar masses. Enhancements in star formation
are strongest at the smallest separations
\citep[$\leq 20$~kpc;][]{patton2013} and closest
mass ratios \citep{ellison2008,scudder2012}, but are detectable at greater than
a factor of $\sim 1.3$ in galaxy pairs separated by up to 80 projected
kpc \citep{scudder2012}, characterized by mass ratios between
0.1 and 10, and velocity differences $\Delta v < 300 \mbox{ km s}^{-1}$.
Elevation in the SFRs of merging pairs relative to isolated
galaxies has also been detected out to $z\sim 1$, both on average, and also
as a function of separation, with closer pairs at $0.5 \leq z \leq 1$ showing
a greater SFR enhancement \citep{wong2011,lin2007}.
To explore the impact of interactions on gas-phase oxygen abundance, \citet{ellison2008} construct
the luminosity-metallicity and mass-metallicity
relations for both paired and isolated galaxies,
demonstrating that galaxies in
pairs are offset by $\sim 0.1$~dex towards lower gas-phase
oxygen abundance at fixed luminosity and by $\sim 0.05$~dex towards
lower metallicity at fixed stellar mass \citep[see also][]{kewley2006,scudder2012}.

The observed differences between the star-forming and chemical abundance properties of
galaxy pairs and their isolated counterparts have been explained using
simulations of galaxy interactions. In \citet{patton2013},
it is shown that simulations of merging galaxies yield
SFR enhancements even out to pair separations
of $\sim 150$~kpc, due to star formation triggered by a preceding
close passage. As for the observed metallicity
differences in galaxy pairs, it is seen that tidal forces
and gravitational torques present during mergers can lead
to inflows of low-metallicity gas into the merging
galaxy nuclei and a corresponding dilution in oxygen
abundance \citep{torrey2012,torrey2017,bustamante2018}.

We used our MOSDEF pairs and control samples to investigate
if differences in SFR and metallicity can be detected in early-stage
mergers relative to isolated galaxies at $1.4 \leq z \leq 2.6$. In Section~\ref{sec:properties-sfrm},
we consider the star-forming main sequence of pairs and control
objects, while, in Section~\ref{sec:properties-mzr}, we explore
potential differences in the MZR.

\subsection{The SFR-M$_*$ Relation}\label{sec:properties-sfrm}
To investigate trends in SFR for interacting and isolated galaxies
at high redshift, we assembled SFR and $M_*$ values for both our
pairs sample and the corresponding control sample over the
redshift range $1.4 \leq z \leq 2.6$. For this
analysis, we considered SFRs derived from dust-corrected
H$\alpha$ and H$\beta$ emission lines (SFR(H$\alpha$)), and also from the same fits to broadband SEDs
that yielded estimates of $M_*$ (SFR(SED)). Prior to comparing the
pairs and control samples, we removed any objects flagged as AGNs based
on X-ray, IR, or optical-emission-line properties\footnote{For this analysis, we
used a simple criterion of $\log(\mbox{NII}\lambda6584/\mbox{H}\alpha) > -0.3$ to flag an object
as an AGN. We did not apply this criterion to galaxies at  $2.95 \leq z \leq 3.80$,
as their spectra lacked coverage of H$\alpha$ and [NII]$\lambda6584$.} \citep{coil2015,azadi2017},
as in such cases both the Balmer emission-line fluxes and the {\it Spitzer}/IRAC
photometric points included in the SED fitting may
be contaminated by radiation associated with the AGN rather
than star formation.

We show the distributions of SFR and $M_*$ for both pairs and control
samples in Figure~\ref{fig:sfrm}, displaying SFR(SED) and SFR(H$\alpha$), respectively,
in the left- and right-hand panels.  Only $\sim 50$\% of the pairs and control samples
in the target redshift range have SFR(H$\alpha$) detections.
The majority of the remaining galaxies have only limits
in SFR(H$\alpha$) because of H$\beta$ or H$\alpha$ non-detections.
For the rest (not plotted), we have no constraints on SFR(H$\alpha$) because
neither H$\alpha$ nor H$\beta$ is detected, or else we lack coverage for
either or both H$\alpha$ and H$\beta$. Given the sample incompleteness,
it is not straightforward to make quantitative inferences
regarding the relative distributions of SFR(H$\alpha$) and $M_*$ for
our pairs and control samples. However, we note that
the distributions of SFR(H$\alpha$) detections and limits for the pairs and control
samples are qualitatively similar.

Both \citet{reddy2015} and \citet{shivaei2016}
have shown that there is general agreement between SFR(SED)
and SFR(H$\alpha$) values for galaxies in the MOSDEF sample.
Therefore, in order to draw more quantitative conclusions regarding
the SFR vs. $M_*$ distributions of interacting and isolated systems,
we use measurements of SFR(SED) vs. $M_*$. We have such
measurements for all galaxies in our pairs and control samples
(with the exception of the unidentified serendip associated with COSMOS-25229), and
therefore incompleteness is not an issue. For a quantitative
comparison, we perform an ordinary least-squares regression fit to
both the pairs and control samples. For the regression analysis,
we restrict the samples to $\log(M_*/M_{\odot})>8$,\footnote{Accordingly,
we do not include the low-mass companion object GOODSN-23418, which is a factor
of 550 lower in mass than its corresponding primary galaxy, GOODSN-23344, and consider
a sample with mass ratios spanning a range comparable to that \citet{ellison2008}.} and exclude
quiescent galaxies obviously offset from the distribution
of star-forming systems (i.e., we do not include
galaxies with $\log(\mbox{SFR(SED)}/M_{\odot}\mbox { yr}^{-1}) \leq -1$ or those with
$\log(M_*/M_{\odot})>11$ and $\log(\mbox{SFR(SED)}/M_{\odot}\mbox { yr}^{-1}) \leq 0.2$).
To obtain error estimates on the best-fit intercept and slope, we perturb individual
$M_*/M_{\odot}$ and SFR(SED) values according to their errors 1000 times, and refit the perturbed
datasets.  The upper and lower confidence bounds we report span 68.3\% of the fits to the perturbed
datasets.  We find for the pairs: 

\begin{multline}\label{eq:sfrmpairs}
\log(\mbox{SFR(SED)})_{\mbox{pairs}}= -4.98^{+1.30}_{-0.90} + \\ 0.60^{+0.09}_{-0.13}\times \log(M_*/M_{\odot})_{\mbox{pairs}}
\end{multline}

\noindent and for the control sample: 

\begin{multline}\label{eq:sfrmcontrol}
\log(\mbox{SFR(SED)})_{\mbox{control}}=-3.80^{+0.45}_{-0.20} + \\ 0.49^{+0.02}_{-0.05}\times \log(M_*/M_{\odot})_{\mbox{control}}
\end{multline}

Accordingly, the best-fit regressions for pairs and control galaxies are
consistent with each other, indicating no elevation in star formation
at fixed mass for galaxies in pairs relative to isolated systems.
If anything, galaxies in pairs are offset towards slightly lower SFR(SED)
at fixed $M_*$, but this offset is not significant. This result holds when
we consider only pairs at separations of $R_{{\rm proj}}\leq 30$~kpc, i.e.,
the separation at which measurable enhancement appears in the local pairs sample in
\citet{ellison2008}. For the $R_{{\rm proj}}\leq 30$~kpc
pair sample, we find very similar best-fit regression parameters to those for the full pairs sample:

\begin{multline}\label{eq:sfrmpairsr30kpc}
\log(\mbox{SFR(SED)})_{\mbox{pairs}}= -4.75^{+1.40}_{-0.90} + \\ 0.58^{+0.10}_{-0.15}\times \log(M_*/M_{\odot})_{\mbox{pairs}}
\end{multline}

We note that the slope and intercept values reported here for the relationship between
SFR(SED) vs. $M_*$ differ from those in \citet{shivaei2015}
(see Table~1 of that paper) and \citet{sanders2018}, mainly due to differences in the methodology
adopted for the regression analysis. However, since our goal here is to 
make a direct comparison between pairs and control galaxies, and since
we use identical regression methodology for those two samples (which we do),
our inferences about their respective SFR(SED) vs. $M_*$ relations are robust.

As another way to quantify the star-forming properties of pairs and control galaxies,
we estimate the median SFR(SED) for the pairs sample used in the regression analysis, and for a subset of the control
sample used for regression analysis that is matched precisely in median stellar mass (i.e., 
$\log(M_*/M_{\odot})_{\rm med}=9.75$). We find $\log(\mbox{SFR(SED)})_{\mbox{pairs,med}}=0.83\pm0.09$ 
for the full sample of pairs, and $\log(\mbox{SFR(SED)})_{\mbox{pairs,med}}=0.84\pm0.09$ for
the subsample with $R_{{\rm proj}}\leq 30$~kpc. For the control sample matched in median stellar mass,
we find $\log(\mbox{SFR(SED)})_{\mbox{control,med}}=0.92\pm0.04$. The difference between pair and control sample
medians is $\Delta SFR_{\rm med}=-0.09\pm0.10$ for the full pairs sample and $\Delta SFR_{\rm med}=-0.08\pm0.11$
for the pairs sample with $R_{{\rm proj}}\leq 30$~kpc. Based on these results, we do not detect
a significant difference in the sample median SFR(SED) values for pairs and control galaxies.
If anything, the pairs sample is offset towards lower median SFR(SED) (but not significantly).

In order to compare with results based on galaxy pairs in SDSS, we make
use of the catalog presented in \citet[][Ellison et al. 2018, private communication]{patton2013,patton2016}. 
We identify a local sample of 830 galaxies in pairs with
$\Delta v\leq 500 \mbox{ km s}^{-1}$, $R_{{\rm proj}}\leq 30$~kpc, and stellar
mass ratios closer than 10:1, and a local control sample 
of 17,835 galaxies having no companions within $\Delta v= 500 \mbox{ km s}^{-1}$ or $R_{{\rm proj}}= 150$~kpc.
In this catalog, SFR is defined as that contained within the SDSS fiber, and is based on
dust-corrected H$\alpha$ emission.  The SDSS pairs and control samples are well matched overall in median and mean stellar mass and redshift,
but the median SFR of the pairs sample is elevated by $\Delta SFR_{\rm med}=0.16\pm0.03$. Comparing the results
for MOSDEF and SDSS pairs with $R_{{\rm proj}}\leq 30$~kpc and $\Delta v\leq 500 \mbox{ km s}^{-1}$
($\Delta SFR_{\rm med}=-0.08\pm0.11$ for MOSDEF pairs and $\Delta SFR_{\rm med}=0.16\pm0.03$ for SDSS pairs),
we find a difference in the SFR enhancement measured at high and low redshift at the $\sim 2\sigma$ level.
A larger sample of high-redshift pairs will reduce the uncertainty on the median properties, and enable
more definitive conclusions.

In addition to considering the global star-forming properties of our pairs sample, 
we also investigate the properties of galaxies in pairs with small mass ratios ($\leq 3$:$1$)
and those with pair separations smaller than 10~kpc, checking
whether such galaxies in particular show enhanced star formation relative to the control sample.
We measure the residuals in SFR(SED) for these two subsamples
of objects around the best-fit linear regression to the full pairs
sample (i.e., $\log(\mbox{SFR(SED)}) - \log(\mbox{SFR(SED)}_{\mbox{fit}}$, where $\log(\mbox{SFR(SED)}_{\mbox{fit}}$
is taken from Equation~\ref{eq:sfrmpairs}), and find no significant vertical offset 
on average for either the small mass ratio or small separation subsample.
Although our high-redshift pairs sample is small, with the associated uncertainties of a small sample size,
our results differ qualitatively 
from the elevation in star formation at fixed mass, and the dependence
of elevated star formation on mass ratio and pair separation
detected for lower-redshift interacting galaxies \citep[e.g.,][]{ellison2008,patton2013,wong2011,lin2007}.
We discuss these differences further in Section~\ref{sec:discussion}.

One final caveat about the results in this section
is that SFR(SED) values we derive are not sensitive to star formation
that is completely obscured at rest-UV and optical wavelengths \citep{chapman2005}. Therefore,
if a higher fraction of the star formation in galaxy pairs is highly
obscured, we will miss it, and miss potential differences in the total bolometric
SFRs of pairs and control galaxies. Deeper far-infrared data are needed to address
this concern.

\begin{figure}[h!]
\includegraphics[width=0.45\textwidth]{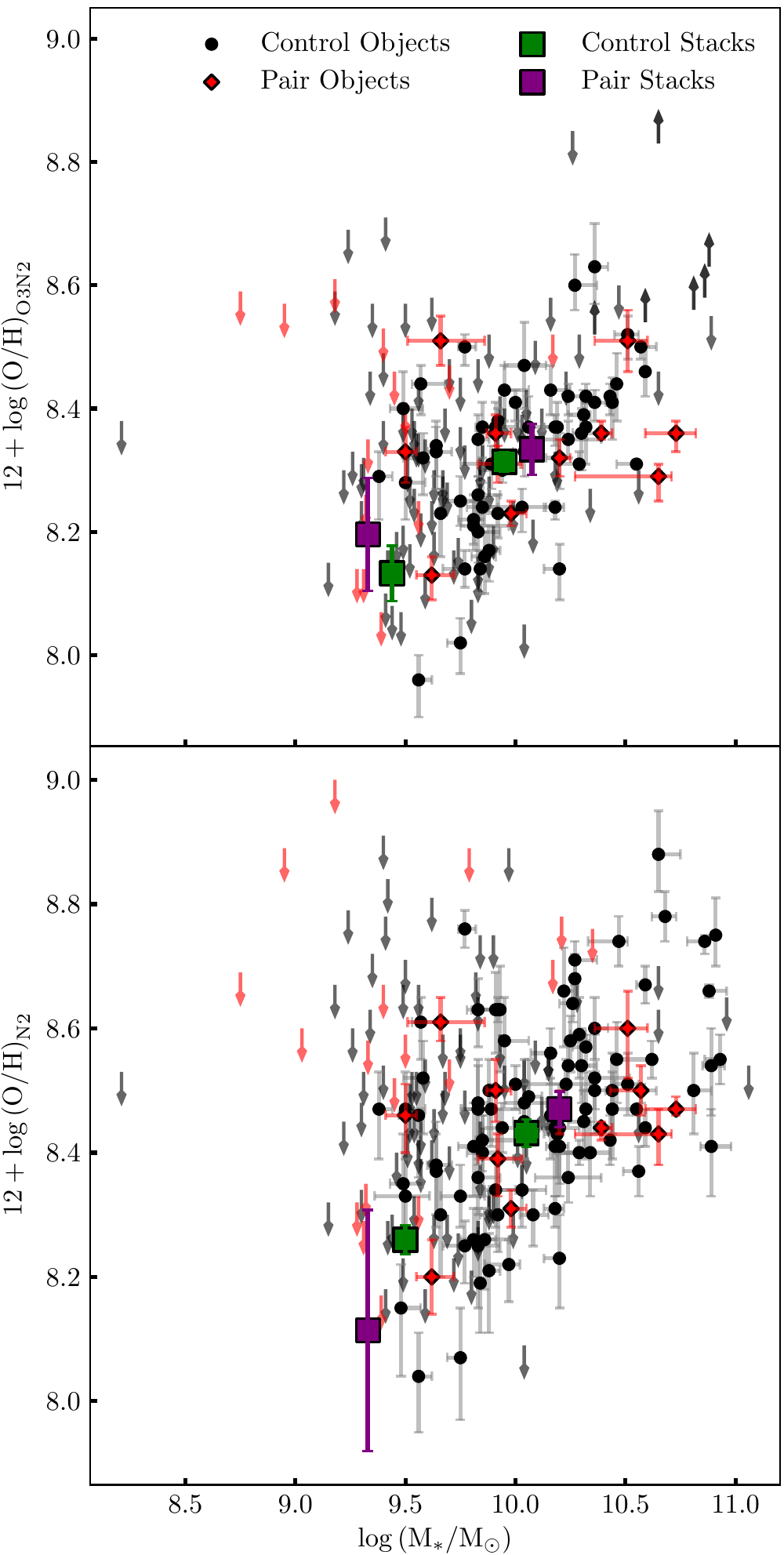}
\caption{MZR for pairs and control objects at $1.4 \leq z \leq 2.6$.
{\bf Top}: In this panel, $12+\log(\mbox{O/H})$ is determined from the $O3N2$
indicator and the calibration of \citet{pettinipagel2004}. Galaxies in pairs are shown
with red symbols, while those in the control sample are indicated in black.
Metallicity limits are indicated as red (grey) arrows for pairs (controls).
In order to include information from individual non-detections, we constructed median
composite spectra for pairs and control galaxies, in two bins of $M_*$ for each sample.
We then measured emission-line ratios from the composite spectra in each stellar
mass bin, and the corresponding $O3N2$ metallicities. Metallicities from the stacked
spectra are shown in purple for the pairs sample, and in green for the control sample.
These stacked measurements indicate no evidence for dilution in metallicity at fixed
$M_*$ for galaxies in the pairs sample.  {\bf Bottom}: In this panel, $12+\log(\mbox{O/H})$
is determined from the $N2$ indicator and the calibration of \citet{pettinipagel2004}. Symbols
are as in the top panel. Measurements from $N2$ composite spectra also indicate
no evidence for metallicity dilution in interacting systems.}
\centering
\label{fig:mzr}
\end{figure}

\subsection{The MZR Relation}\label{sec:properties-mzr}
We also analyzed the distribution of pairs and control galaxies in the
space of metallicity and mass, considering the MZR
for these two samples. In this analysis, we used two different metallicity
indicators, $N2$ and $O3N2$ \citep{pettinipagel2004}, to check whether the results depended
on the method for estimating oxygen abundance. We again restricted this analysis to
the redshift range $1.4 \leq z \leq 2.6$, over which H$\beta$, [OIII]$\lambda 5007$, 
H$\alpha$ and [NII]$\lambda 6584$ emission lines are all accessible in ground-based
spectroscopy, and removed AGNs as in Section~\ref{sec:properties-sfrm}.
We show the distributions of $12+\log(\mbox{O/H})$ and $M_*$ in Figure~\ref{fig:mzr}
for both pairs and control galaxies. As in the case of SFR(H$\alpha$),
our galaxy samples have roughly equal numbers of detections and limits,
mainly due to non-detections in H$\beta$ or  [NII]$\lambda 6584$.

In order to quantify the average trends in mass and metallicity and
incorporate information from galaxies with individual limits,
we construct median composite spectra in two bins of stellar mass
for both pairs and control samples, following the methodology described
in \citet{sanders2018}. The only requirements for inclusion in these stacks
are coverage of H$\alpha$ and [NII]$\lambda 6584$ in the case of the $N2$ stacks,
and additional coverage of [OIII]$\lambda 5007$ and H$\beta$ for the $O3N2$ stacks, and a
detection of H$\alpha$ emission at $\geq 3\sigma$. Accordingly,
the sample of objects included in the $O3N2$
stacks is a subset of the objects in the $N2$ stacks, given the additional requirement
of [OIII]$\lambda 5007$ and H$\beta$ coverage. For composite measurements
of [NII]$\lambda 6584$/H$\alpha$, we used $H$-band spectra for galaxies at $1.4 \leq z \leq 1.7$,
and $K$-band spectra for those at $1.9 \leq z \leq 2.7$. For the corresponding
measurements of [OIII]$\lambda 5007$/H$\beta$, we used $J$-band spectra
for galaxies at $1.4 \leq z \leq 1.7$ and $H$-band spectra for those at $1.9 \leq z \leq 2.7$.
H$\alpha$ and H$\beta$ emission-line measurements from the composite spectra were corrected for
underlying stellar Balmer absorption based on the median absorption strength for individual
galaxies going into the stack, as estimated from stellar population synthesis models
to broadband photometry. Dust extinction corrections were not applied to the composite
spectra, given the close proximity of [NII]$\lambda 6584$ and H$\alpha$, and that
of [OIII]$\lambda 5007$ and H$\beta$.  The median $M_*$ in each stellar mass bin
and $12+\log(\mbox{O/H})$ estimated from the corresponding composite spectrum are
are shown in Figure~\ref{fig:mzr}, along with the individual
measurements and limits. It is clear that the pairs and control galaxies follow very similar trends
in $12+\log(\mbox{O/H})$, regardless of whether $N2$ or $O3N2$ is considered. The same
results hold when we consider the subset of MOSDEF pairs at $R_{{\rm proj}}\leq 30$~kpc.
We recall here that local pairs show a depression in metallicity at fixed luminosity
and stellar mass relative to isolated systems  \citep[e.g.,][]{ellison2008,scudder2012}, in apparent contrast to our own 
preliminary high-redshift results. However, the uncertainty on the metallicities estimated from the pairs composite spectra
is equal to or larger than the observed $z\sim 0$ depression in metallicity ($\sim 0.05$~dex).  
We therefore cannot rule out a $\leq 0.05$~dex offset in metallicity between interacting pairs and isolated galaxies at fixed 
$M_*$.

\section{Discussion and Future Directions}\label{sec:discussion}
We have demonstrated that merging galaxy pairs at $1.4 \leq z \leq 2.6$
are not characterized by elevated SFRs or significantly diluted metallicities
relative to isolated systems of the same stellar mass. Although
we will require a larger pairs sample to place these results on a more secure statistical
footing, it is worthwhile to consider the implications of the suggested trends, which run
contrary to the patterns uncovered among interacting galaxies at lower redshift.
As we discuss below, there is
a theoretical basis for the apparent evolution in the differential
star-forming properties of merging systems. Here we compare our results with
other differential studies of merging pairs at $z>1$, consider the predictions
from simulations, and discuss evidence for AGN activity in our merging
systems. We conclude by listing some promising future directions
for the study of interacting galaxies at high redshift.

\subsection{Comparisons with Other Observational Work}\label{sec:discussion-comparison}
Our study represents the first controlled differential comparison of 
SFRs in interacting and isolated galaxies at $z>1$. However,
\citet{divoy2014} previously considered the question of the relationship
between metallicity and small-scale environment.
These authors analyzed a sample of 49 star-forming
galaxies at $z\sim 0.9-1.8$ with VLT/SINFONI integral field unit
emission-line maps and associated $N2$ metallicity measurements.
In this sample, 12 systems are identified as ``interacting"
based on the presence of a companion within $\Delta v=500 \mbox{ km s}^{-1}$
and $R_{{\rm proj}}=30 h^{-1}$~kpc \citep{lopezsanjuan2013}.
\citet{divoy2014} find a measurable depression 
in median metallicity of at least 0.13~dex for their ``interacting" sample
relative to a control sample of 37 isolated galaxies at similar median stellar mass.
This depression is significantly larger than what is observed in the local universe
\citep[$\leq 0.05$~dex;][]{ellison2008,scudder2012}, and in contrast to the lack of 
significant offset in metallicity that we find for our pairs relative to the control sample. In order to obtain
a more robust result at high redshift, we require a significantly larger
sample of galaxy pairs with metallicity measurements, and deeper spectroscopy
sufficient to detect H$\beta$ and [NII]$\lambda 6584$ for a larger fraction
of the sample in order to reduce the number of metallicity upper limits.

\subsection{Expectations From Simulations}\label{sec:discussion-simulations}
The lack of enhancement in SFR observed for the MOSDEF pairs sample is
consistent with recent predictions from numerical simulations of galaxy formation.
\citet{fensch2017} run a suite of pc-scale galaxy merger simulations, representing low-redshift
galaxies with gas fractions of 10\%, while $z=2$ galaxies are simulated
with gas fractions of 60\%. For the same orbital parameters, it is found that
the gas-poor merger simulations approximating local galaxies
feature boosts in the SFR of the merging galaxies of an order of magnitude or more over an extended
period of hundreds of Myr prior to coalescence, while the gas-rich
simulations approximating $z=2$ show only mild increases in SFR, and only
at coalescence. Based on a suite of binary galaxy merger simulations with identical
orbital parameters but nine different initial gas fractions ranging from $M_{gas}/M_* = 0.04$ to $1.78$,
\citet{scudder2015} similarly find that the enhancement in SFR during the merger is anti-correlated
with initial gas fraction. \citet{fensch2017} attribute the difference in the evolution of the SFR during
the merger to differences in the increase of both central gas inflows and compressive
turbulence in the ISM during mergers at $z=2$ and $z=0$. Both gas-rich ($z=2$)
and gas-poor ($z=0$) merger simulations reach similar peak central gas inflow rates fueling star formation,
but, given that the  gas-rich simulations start off with pre-merger baseline gas inflow rates that are
an order of magnitude higher than those in the gas-poor simulations, the enhancement in gas inflow
and corresponding SFR is significantly weaker. \citet{fensch2017} then
attribute the lack of increase in ISM turbulence during $z=2$ mergers to both an
ISM velocity dispersion that is higher pre-merger and harder to additionally stir-up
\citep{wisnioski2015}, and a clumpy ISM architecture with an associated tidal field that also suppresses an
increase in turbulence \citep{genzel2008}. These simulations do not incorporate cosmological
accretion or the more compact nature of $z=2$ galaxies relative to those at $z=0$ of the
same mass, though the authors argue that differences in gas fraction and ISM structure
are the dominant effects underlying the observed difference in SFR evolution for high-redshift
mergers.  Analysis of the lower-resolution Horizon-AGN cosmological hydrodynamical simulation
by \citet{martin2017} also predicts that the enhancement in SFR
at fixed mass due to mergers is most pronounced at low redshift,
and undetectable at $z>1.5$, when the ambient pre-merger level of star formation
is an order of magnitude higher than in the local universe.

There is also recent theoretical work focusing on the evolution of ISM metallicity during
galaxy mergers.  Based on a sample of 70 gas-rich mergers traced at $z<1.5$
in the Auriga cosmological simulation, \citet{bustamante2018}
find that a period of metallicity dilution typically occurs during merger
events, reaching a magnitude of $\Delta Z=-0.1$~dex for major mergers at projected
separations $<10$~kpc, and at least a few hundreths of a dex depression for both major
and minor mergers at separations of $<30$~kpc. These results are roughly
consistent with observations of $z\sim 0$ merging pairs by \citet[e.g.,][]{scudder2012}.
However, it is also worth noting that the pre-coalescence depressions in metallicity recovered
in these simulations are consistent with those inferred from the fundamental
metallicity relation \citep[FMR;][]{mannucci2010,ellison2008fmr}. Specifically, given
the relation among $M_*$, SFR, and gas-phase oxygen abundance in the FMR, the metallicity
during the mergers is consistent with predictions from the FMR, given the evolution in
SFR and $M_*$ of the merging galaxies.
Accordingly, given that our merging pairs show no offset from the 
control sample in the SFR vs. $M_*$ relation, and given that these pairs represent a pre-coalescence
phase, the expectation from the simulation results of \citet{bustamante2018} is
that pairs in MOSDEF should simply follow the same
relationship among $M_*$, SFR, and metallicity that is inferred at $z\sim 2$
for the MOSDEF sample as a whole \citep{sanders2018}. Furthermore,
given the size of the error bars on our median stacked metallicities, we are 
not sensitive to differences of $\leq 0.05$~dex as observed
by \citet{ellison2008} and \citet{scudder2012}, and predicted for all but the most extreme merger
events in \citet{bustamante2018}.  

Finally, we note that \citet{torrey2017}
have also found evidence for metallicity dilution during mergers in the 
IllustrisTNG cosmological simulation, but the detailed example analyzed
in that work consisted of a merger at $z\sim 0$. It is not clear
how the analogous results would differ at $z\sim 2$. High-resolution
``$z=2$" simulations like those of \citet{fensch2017} are required to address
the question of metallicity dilution in high-redshift mergers. For robust conclusions,
such simulations must also include
cosmological gas accretion and track metal enrichment as well as star formation.

\subsection{The AGN Fraction in MOSDEF Pairs}\label{sec:discussion-agn}
Our samples of galaxies in pairs are indistinguishable from their isolated
counterparts in terms of star formation and metallicity at fixed
stellar mass. The extensive multi-wavelength data available for MOSDEF
targets also enables AGN classifications on the basis of X-ray luminosity,
infrared colors, and/or rest-optical emission line ratios \citep{coil2015,azadi2017}.
Given the connection between galaxy mergers and black hole fueling inferred from numerical
simulations \citep[e.g.,][]{hopkins2006,dimatteo2005},
much observational work has been devoted to exploring the link between galaxy interactions
and AGN activity, with mixed results. For example, \citet{ellison2011}
demonstrate an enhancement in AGN fraction among galaxy close pairs at $z\sim 0$,
with up to a factor of 2.5 enhancement in AGN fraction for pairs with projected
separations less than 40~kpc. Along the same lines, \citet{goulding2018} use machine-learning classifications
of merging systems at $z\sim 0-0.9$, finding an AGN fraction (as estimated by
WISE infrared colors) elevated by a factor of
$\sim 2-7$ among interacting relative to non-interacting systems.
Many other investigations have focused instead on the merger fraction among AGNs
(as opposed to the AGN fraction among mergers), finding that AGNs do not show a significantly enhanced
fraction of mergers relative to their inactive counterparts \citep[e.g.,][]{kocevski2012,gabor2009,hewlett2017}.
Exploring the merger fraction among AGNs and non-AGNs provides constraints on how important interactions
are for driving AGN activity relative to other more secular processes \citep{bournaud2011}. Given
the design of our experiment and the incompleteness of our pairs sample (see Section~\ref{sec:control}),
we are not in the position to quantify the differential merger fraction among
AGNs and non-AGNs. However, we can obtain complete estimates of the AGN fraction among
merging and isolated galaxies, which indicates how likely it is for mergers
to trigger AGN activity.

Based on the X-ray, infrared, and rest-optical emission-line AGN
classifications for MOSDEF galaxies, we find that in our full pairs sample the fraction
of galaxies satisfying at least one of these AGN criteria
is 10 out of 59 ($16.9\% \pm 5.3$\%, with the error based on simple Poisson statistics).
Focusing on the $1.4 \leq z \leq 2.6$ subsample for which we compared SFRs and metallicities
with their isolated counterparts, we find 9 out of 47 ($19.1\% \pm 6.3$\%) are classified as AGNs.
We also note that 7 out of 36 ($19.4\% \pm 7.3$\%) of these systems with $R_{{\rm proj}}\leq 30$~kpc --
i.e., the same fraction -- satisfy at least one of the AGN criteria.
The corresponding AGN fractions for the full and $1.4 \leq z \leq 2.6$ control subsamples
are 42 out of 372 ($11.3\% \pm1.7$\%) and 32 out of 242 ($13.2\%\pm2.3$\%). Accordingly, we find
a higher AGN fraction among pairs relative to controls, with a factor of at least
$\sim 1.5$ enhancement in the pair AGN fraction at $1.4 \leq z \leq 2.6$. However,
given the small sample size of the pairs, this difference is not highly significant.
Increasing the sample size of pairs and control galaxies by an order of magnitude will 
enable robust statistics on the differential AGN fraction between pairs and control
galaxies.

\subsection{Future Work}\label{sec:discussion-future}
Our analysis comprises an early step in characterizing the properties
of merging galaxies at high redshift. With only 30 pairs in total,
24 of which we analyze in detail along with a carefully-defined
control sample of isolated galaxies, the statistical power of our sample 
is limited. For example, while \citet{ellison2008} and \citet{scudder2012} divided their samples
of almost 2000 merging pairs into multiple bins of projected
separation and mass ratio to explore how the physical properties of merging galaxies depend on each of these merger
characteristics, our current small sample size precludes such division.
In addition to obtaining a significantly larger sample of merging pairs,
we must also obtain significantly deeper rest-optical emission-line spectra,
so that the fraction of limits in SFR(H$\alpha$) and $12+\log(\mbox{O/H})$
is reduced to a negligible contribution across a sufficiently wide mass range, and we can therefore analyze
the distributions of individual merging and isolated
galaxies in the spaces of SFR(H$\alpha$) vs. $M_*$ and $12+\log(\mbox{O/H})$
vs. $M_*$. 

We also need to apply an effective technique for identifying
later-stage mergers closer to coalescence, when the effects of 
SFR enhancement and metallicity dilution are predicted to be strongest
\citep{bustamante2018}. At low redshift, non-parametric morphological measures
are commonly used to identify such mergers \citep{lotz2004}, but \citet{abruzzo2018} demonstrate
with mock observations of cosmological simulations that these same
merger morphological classifications at $z\sim 2$ are significantly contaminated 
by non-merging galaxies and should not be applied to {\it HST} images
of high-redshift galaxies. Another route to studying later-stage mergers would
consist of building up a significantly larger (i.e., two orders
of magnitude) sample of spectroscopically-confirmed pairs
with $R_{{\rm proj}}<10$~kpc pairs and $\Delta v \leq 500 \mbox{ km s}^{-1}$.
With larger galaxy pairs samples accompanied by complete sets of emission-line detections
and a larger range of merger stages, we will truly be able to test models
of galaxy mergers in the early universe.

\begin{deluxetable*}{lrccrccrrr}
\tablecolumns{8}
\tablewidth{7in}
\tablecaption{Observed Properties of MOSDEF Pairs\label{tab:basic}}
\tablehead{
\colhead{Field} & \colhead{ID\tablenotemark{a}} & \colhead{$z_{\rm{MOSFIRE}}$\tablenotemark{b}} & \colhead{$m_{\rm{F160W}}$\tablenotemark{c}} & \colhead{ID\tablenotemark{a}}&  \colhead{$z_{\rm{MOSFIRE}}$\tablenotemark{b}} & \colhead{$m_{\rm{F160W}}$\tablenotemark{c}} &\colhead{ $R_{\rm{proj}}$\tablenotemark{d}} & \colhead{$\Delta v$\tablenotemark{e}} & \colhead{$M_{*{\rm Larger}}/M_{*{\rm Smaller}}$\tablenotemark{f}}\\
\colhead{} & \colhead{(Primary)} & \colhead{} & \colhead{} & \colhead{(Serendip)} &  \colhead{} & \colhead{} &\colhead{(kpc)} & \colhead{($\mbox{km s}^{-1}$)} & \\
}
\startdata
\\
\multicolumn{9}{c}{$1.37\leq z \leq 1.70$} \\
\hline
COSMOS   & 8381      &    1.4049         & 21.21      & 8490                 &  1.4052               & 23.55      & 38.4                               & 37     & 7.6                       \\
COSMOS   & 19325     &    1.6013         & 23.86      & 19292                &  1.6007               & 24.16      & 11.1                               & 74     & 1.9                       \\
GOODSN  & 10044     &    1.6006         & 23.14      & 10041                &  1.6039               & 25.33      & 53.1                               & 380     & 11.0                      \\
AEGIS    & 35056     &    1.6534         & 23.81      & 35075                &  1.6530               & 24.27      & 5.7                                & 38     & 1.1                       \\
AEGIS    & 39567     &    1.5805         & 21.95      & 39897                &  1.5806               & 23.89      & 10.7                               & 17     & 7.8                        \\
AEGIS    & 16339     &    1.5291         & 21.13      & 16026                &  1.5320               & 21.69      & 4.8                                & 338    & 1.8                        \\
AEGIS    & 3237      &    1.6667         & 22.38      & 3478                 &  1.6657               & 22.66      & 6.8                                & 113    & 8.1                        \\
\hline\\
\multicolumn{9}{c}{$1.90\leq z \leq 2.61$} \\
\hline
COSMOS   &   1740    & 2.2999            & 23.89      & 1795                 & 2.2985                & 26.04      & 39.4                               & 121         & 4.4                   \\
COSMOS   &   10719   & 2.2465            & 23.61      & 10766                & 2.2505                & 25.07      & 9.9                               & 368          & 6.8                  \\
COSMOS   &   471     & 2.0731            & 23.52      & 532                  & 2.0678                & 24.55      & 52.5                               & 521         & 3.8                   \\
COSMOS   &   21492   & 2.4721            & 20.99      & 21598                & 2.4786                & 24.06      & 20.5                               & 561         & 7.8                   \\
COSMOS   &   14849   & 1.9265            & 22.08      & 14968                & 1.9233                & 21.99      & 11.8                               & 330         & 3.5                   \\
COSMOS   &   25229   & 2.1813            & 22.79      & $-$9999\tablenotemark{g}   & 2.1826           & \nodata\tablenotemark{g}     & 12.8              & 130       & \nodata                    \\
COSMOS   &   7433    & 2.1662            & 24.03      & 7417                 & 2.1661                & 24.70      & 8.0                                & 8           & 2.0                   \\
GOODSN  &   17748   & 2.2325            & 24.20      & 17714                & 2.2349                & 23.67      & 10.2                               & 221          & 1.5                  \\
GOODSN  &   24825   & 2.3347            & 23.63      & 25017                & 2.3359                & 23.47      & 29.7                               & 110          & 1.3                  \\
GOODSN  &   23344\tablenotemark{h}   & 2.4839            & 23.25      & 23418                & 2.4889                & 26.39      & 45.1              & 433          & 550.0                 \\
GOODSN  &   23344\tablenotemark{h}   & 2.4839            & 23.25      & 23339                & 2.4828                & 23.53      & 12.5              & 92           & 7.6                  \\
GOODSN  &   23869   & 2.2438            & 23.15      & 24074                & 2.2433                & 24.37      & 50.2                               & 46           & 5.0                  \\
GOODSN  &   12302   & 2.2756            & 21.97      & 12172                & 2.2754                & 25.06      & 10.5                               & 16           & 17.0                 \\
AEGIS    &   18543   & 2.1387            & 23.15      & 18454                & 2.1384                & 24.23      & 12.4                               & 33          & 2.0                   \\
AEGIS    &   31108   & 2.3547            & 22.68      & 31317                & 2.3570                & 23.30      & 24.3                               & 205         & 5.4                   \\
AEGIS    &   36050   & 2.5285            & 23.31      & 36180                & 2.5325                & 23.07      & 17.7                               & 339         & 3.9                   \\
AEGIS    &   29114   & 2.3519            & 24.46      & 29045                & 2.3500                & 24.94      & 8.6                                & 175         & 2.3                   \\
\hline\\
\multicolumn{9}{c}{$2.95\leq z \leq 3.80$} \\ 
\hline
COSMOS   & 23183     & 3.1211            & 24.45      & 23192                & 3.1227                & 24.87      & 9.2                                & 123         & 1.6                   \\
COSMOS   & 24579     & 3.2521            & 25.04      & 24596                & 3.2566                & 25.84      & 11.6                               & 315         & 4.4                   \\
COSMOS   & 2360      & 3.0338            & 24.34      & 2344                 & 3.0344                & 25.08      & 16.5                               & 39          & 2.2                   \\
GOODSN  & 28202     & 3.2325            & 23.87      & 28209                & 3.2336                & 23.78      & 5.4                                & 76           & 1.1                  \\
GOODSN  & 15694     & 3.3233            & 24.71      & 15566                & 3.3203                & 26.68      & 13.1                               & 207          & 9.5                  \\
AEGIS    & 30847     & 3.4349            & 23.51      & 30691                & 3.4431                & 24.92      & 16.9                               & 553         & 3.1                   
\enddata
\tablenotetext{a}{Galaxy ID in the 3D-HST v4.1 catalogs \citep{momcheva2016}.}
\tablenotetext{b}{Emission-line redshift measured from MOSFIRE spectroscopy as part of the MOSDEF survey.}
\tablenotetext{c}{{\it HST}/WFC3 F160W magnitude on the AB system.}
\tablenotetext{d}{Projected separation in kpc, based on the angular separation between the pair galaxies and the redshift of the primary galaxy.}
\tablenotetext{e}{Velocity separation in $\mbox{km s}^{-1}$, based MOSFIRE systemic redshifts of the primary and serendip galaxies.}
\tablenotetext{f}{Stellar mass ratio between more passive and less massive pair member, irrespective of which galaxy is the ``primary" and which is the ``serendip."}
\tablenotetext{g}{Although we identify an object in the F160W image at the location corresponding to where our MOSFIRE serendip spectrum was extracted, there is no 3D-HST catalog entry for this galaxy. We list the ID for this serendip as ``$-$9999" and lack a robust F160W magnitude. We 
measure its coordinates directly from the F160W image, from which we determine the projected separation for this pair.}
\tablenotetext{h}{Primary galaxy with two serendip companions. The primary galaxy is listed twice, once for each serendip.}
\tablecomments{The three columns immediately to the right of the primary galaxy ID list the properties of the primary galaxy in each pair, while the corresponding columns immediately
to the right of the serendip ID refer to the serendip. The final two columns refer to the properties of the pair.}
\end{deluxetable*}

\begin{deluxetable*}{lrlrcrrr}
\tablecolumns{8}
\tablewidth{7in}
\tablecaption{Physical Properties of MOSDEF Pairs\label{tab:physical}}
\tablehead{
\colhead{Field} & \colhead{ID} & \colhead{Primary/} & \colhead{$\log(M_*)$\tablenotemark{a}} & \colhead{$\log(\mbox{SFR(SED)})$\tablenotemark{b}} & \colhead{$\log(\mbox{SFR(H}\alpha))$\tablenotemark{c}} &  \multicolumn{2}{c}{$12+\log(\mbox{O/H})$\tablenotemark{d}}  \\
\colhead{} & \colhead{} & \colhead{Serendip} & \colhead{$(M_{\odot})$} & \colhead{$(M_{\odot} \mbox{ yr}^{-1})$} & \colhead{$(M_{\odot} \mbox{ yr}^{-1})$} & \colhead{($O3N2$)} & \colhead{($N2$)} 
}
\startdata
\multicolumn{8}{c}{$1.37\leq z \leq 1.70$} \\
\hline
COSMOS\tablenotemark{e} & 8381 & primary & 10.94 & 0.60 & \nodata & $<$8.74 & $<$8.88 \\
COSMOS & 8490 & serendip & 10.06 & 0.00 & \nodata & \nodata & \nodata \\
COSMOS & 19325 & primary & 9.45 & 0.76 & 0.40 & $<$8.46 & $<$8.52 \\
COSMOS & 19292 & serendip & 9.18 & 0.56 & $>$0.91 & $<$8.61 & $<$9.00 \\
GOODSN & 10044 & primary & 9.79 & 0.83 & $>$0.15 & \nodata & $<$8.89 \\
GOODSN & 10041 & serendip & 8.75 & -0.21 & $>$0.00 & $<$8.59 & $<$8.69 \\
AEGIS & 35056 & primary & 9.31 & 0.68 & $>$1.21 & $<$8.14 & $<$8.29 \\
AEGIS & 35075 & serendip & 9.28 & 0.32 & 1.13 & $<$8.14 & $<$8.32 \\
AEGIS & 39567 & primary & 10.39 & 1.43 & 1.96 & 8.36 & 8.44 \\
AEGIS & 39897 & serendip & 9.50 & 0.83 & 0.70 & 8.33 & 8.46 \\
AEGIS\tablenotemark{e} & 16339 & primary & 10.55 & 2.25 & 2.36 & 8.61 & 8.58 \\
AEGIS\tablenotemark{e} & 16026 & serendip & 10.83 & 1.87 & 1.70 & \nodata & 8.71 \\
AEGIS & 3237 & primary & 10.57 & 0.88 & $>$1.15 & \nodata & 8.50 \\
AEGIS & 3478 & serendip & 9.66 & 1.75 & 0.95 & 8.51 & 8.61 \\
\hline
\multicolumn{8}{c}{$1.90\leq z \leq 2.61$} \\
\hline
COSMOS & 1740 & primary & 9.91 & 1.05 & 1.40 & 8.36 & 8.50 \\
COSMOS & 1795 & serendip & 9.27 & -0.25 & $<$1.20 & \nodata & \nodata \\
COSMOS & 10719 & primary & 11.04 & 0.88 & \nodata & \nodata & \nodata \\
COSMOS & 10766 & serendip & 10.21 & 0.69 & $>$0.67 & \nodata & $<$8.78 \\
COSMOS\tablenotemark{e} & 471 & primary & 9.61 & 0.94 & \nodata & \nodata & 8.33 \\
COSMOS & 532 & serendip & 9.03 & 1.55 & \nodata & \nodata & $<$8.60 \\
COSMOS\tablenotemark{e} & 21492 & primary & 11.06 & 2.39 & 2.21 & 8.39 & 8.63 \\
COSMOS & 21598 & serendip & 10.17 & 0.94 & $>$0.37 & $<$8.52 & $<$8.71 \\
COSMOS & 14849 & primary & 10.35 & 1.92 & \nodata & \nodata & $<$8.76 \\
COSMOS\tablenotemark{e} & 14968 & serendip & 9.80 & 1.97 & \nodata & \nodata & 8.70 \\
COSMOS & 25229 & primary & 9.98 & 1.31 & 1.42 & 8.23 & 8.31 \\
COSMOS & $-$9999 & serendip &  \nodata & \nodata & \nodata & \nodata & \nodata \\
COSMOS & 7433 & primary & 9.70 & 0.84 & $>$0.83 & $<$8.47 & $<$8.55 \\
COSMOS & 7417 & serendip & 9.40 & 0.79 & $>$0.20 & $<$8.53 & $<$8.67 \\
GOODSN & 17748 & primary & 9.56 & 0.60 & 0.94 & $<$8.25 & $<$8.33 \\
GOODSN & 17714 & serendip & 9.39 & 0.72 & 2.07 & $<$8.07 & $<$8.17 \\
GOODSN & 24825 & primary & 9.75 & 0.43 & \nodata & \nodata & \nodata \\
GOODSN\tablenotemark{e} & 25017 & serendip & 9.65 & 0.70 & 1.36 & \nodata & \nodata \\
GOODSN & 23344 & primary & 10.65 & 1.78 & 2.34 & 8.29 & 8.43 \\
GOODSN & 23418 & serendip & 7.91 & 0.66 & $<$1.37 & \nodata & \nodata \\
GOODSN & 23344 & primary & 10.65 & 1.78 & 2.34 & 8.29 & 8.43 \\
GOODSN & 23339 & serendip & 9.77 & 1.33 & $<$0.95 & \nodata & \nodata \\
GOODSN & 23869 & primary & 10.20 & 1.53 & 1.99 & 8.32 & 8.47 \\
GOODSN & 24074 & serendip & 9.50 & 0.35 & $>$0.45 & $<$8.41 & $<$8.59 \\
GOODSN\tablenotemark{e} & 12302 & primary & 10.92 & 2.11 & 1.99 & 8.56 & 8.68 \\
GOODSN\tablenotemark{e} & 12172 & serendip & 9.69 & 0.32 & 1.40 & 8.36 & 8.45 \\
AEGIS & 18543 & primary & 9.62 & 0.99 & 1.33 & 8.13 & 8.20 \\
AEGIS & 18454 & serendip & 9.33 & 0.37 & $>$1.99 & $<$8.35 & $<$8.58 \\
AEGIS & 31108 & primary & 10.73 & 1.91 & 2.17 & 8.36 & 8.47 \\
AEGIS & 31317 & serendip & 10.00 & 0.68 & \nodata & \nodata & \nodata \\
AEGIS & 36050 & primary & 9.92 & 1.30 & 1.37 & 8.31 & 8.39 \\
AEGIS & 36180 & serendip & 10.51 & 1.72 & 1.47 & 8.51 & 8.60 \\
AEGIS & 29114 & primary & 9.32 & 0.36 & 0.62 & $<$8.26 & $<$8.35 \\
AEGIS & 29045 & serendip & 8.95 & 0.51 & 0.24 & $<$8.57 & $<$8.89 \\
\hline
\multicolumn{8}{c}{$2.95\leq z \leq 3.80$} \\
\hline
COSMOS & 23183 & primary & 10.06 & 1.82 & \nodata & \nodata & \nodata \\
COSMOS & 23192 & serendip & 10.27 & 1.60 & \nodata & \nodata & \nodata \\
COSMOS & 24579 & primary & 9.42 & 0.80 & \nodata & \nodata & \nodata \\
COSMOS & 24596 & serendip & 8.78 & 0.54 & \nodata & \nodata & \nodata \\
COSMOS & 2360 & primary & 9.73 & 1.06 & \nodata & \nodata & \nodata \\
COSMOS & 2344 & serendip & 9.39 & 0.90 & \nodata & \nodata & \nodata \\
GOODSN\tablenotemark{e} & 28202 & primary & 10.17 & 1.50 & \nodata & \nodata & \nodata \\
GOODSN & 28209 & serendip & 10.12 & 1.45 & \nodata & \nodata & \nodata \\
GOODSN & 15694 & primary & 9.81 & 1.14 & \nodata & \nodata & \nodata \\
GOODSN & 15566 & serendip & 8.83 & 0.34 & \nodata & \nodata & \nodata \\
AEGIS & 30847 & primary & 9.30 & 2.04 & \nodata & \nodata & \nodata \\
AEGIS & 30691 & serendip & 8.81 & 1.56 & \nodata & \nodata & \nodata
\enddata
\tablenotetext{a}{Log stellar mass, in units of $M_{\odot}$.}
\tablenotetext{b}{Log of SFR derived from the best-fit population synthesis model, corrected for dust extinction, in units of $M_{\odot} \mbox{ yr}^{-1}$.}
\tablenotetext{c}{Log of SFR derived from H$\alpha$ luminosity, corrected for dust extinction, in units of  $M_{\odot} \mbox{ yr}^{-1}$. An entry of ``\nodata" indicates galaxies lacking
coverage of either H$\alpha$ or H$\beta$, or with limits in both H$\alpha$ and H$\beta$. SFRs for galaxies with H$\alpha$ detections and H$\beta$ non-detections
are indicated as lower limits, while those for galaxies with H$\alpha$ non-detections and H$\beta$ detections are shown with upper limits.}
\tablenotetext{d}{Gas-phase oxygen abundance. The column labeled $O3N2$ lists oxygen abundances based on the $O3N2$ indicator,
while the column labeled $N2$ contains oxygen abundances based on the $N2$ indicator \citep{pettinipagel2004}. A value of ``\nodata" indicates
galaxies lacking coverage of at least one of the required emission lines, or with enough non-detections to prevent derivation of a meaningful
limit. Accordingly, we do not report oxygen abundances for galaxies in the highest-redshift bin.}
\tablenotetext{e}{Galaxy identified as an AGN based on X-ray, infrared, or rest-optical emission-line properties. Stellar mass, SFR, and metallicity
values are not plotted or included in differential comparisons.}
\end{deluxetable*}

\acknowledgements We thank the referee for an extremely constructive and thorough report. We are deeply grateful
to Sara Ellison and David Patton for sharing their SDSS catalog of local star-forming galaxies and their closest companions. We acknowledge support from NSF AAG grants AST-1312780, 1312547, 1312764, and 1313171, archival grant AR-13907 provided by NASA through the Space Telescope Science Institute, and grant NNX16AF54G from the NASA ADAP program. This work was supported by the NSF under  NSF REU Grant No. PHY-1460055, along with support from the Department of Physics \& Astronomy of the University of California, Los Angeles. R.L.S. was supported by a UCLA Graduate Division Dissertation Year Fellowship. We also acknowledge a NASA contract supporting the ``WFIRST Extragalactic Potential Observations (EXPO) Science Investigation Team" (15-WFIRST15-0004), administered by GSFC. We also acknowledge the 
3D-HST collaboration, who provided us with spectroscopic and photometric catalogs used to select MOSDEF targets and derive
stellar population parameters. We wish to extend special thanks to those of Hawaiian ancestry on
whose sacred mountain we are privileged to be guests. Without their generous hospitality, most
of the observations presented herein would not have been possible.

\bibliography{merger}
\bibliographystyle{apj}

\end{document}